\documentclass[11pt,letterpaper]{amsart}
\usepackage{multicol,amssymb,amsmath,amsbsy,amsthm,graphicx}
\usepackage[latin1]{inputenc}

\theoremstyle{plain}
\newtheorem{theo}{Theorem}[section]
\newtheorem{lem}[theo]{Lemma}
\newtheorem{co}[theo]{Corollary}
\newtheorem{prop}[theo]{Proposition}

\theoremstyle{definition}
\newtheorem{defn}[theo]{Definition}

\newcommand{\supp}{\text{\rm{supp}}}
\newcommand{\vol}{\text{\rm{Vol}}}
\newcommand{\card}{\text{\rm{card}}}
\newcommand{\diam}{\text{\rm{diam}}}
\newcommand{\AEM}{\text{\rm{AEM}}}
\newcommand{\CPD}{\text{\rm{CPD}}}

\begin{document}

\begin{center}
{\bf \LARGE{Asymptotics of Greedy Energy Points}}\\

\vspace{0.5cm}

{A. L\'opez Garc\'ia}\footnotemark\footnotetext{The results of
this paper form a part of this author's Ph.D. dissertation at
Vanderbilt University.} \hspace{1cm} {E. B.
Saff}\,\footnotemark\footnotetext{The research of this author was
supported, in part, by National Science Foundation grants
DMS-0603828 and DMS-0808093.}
\end{center}

{\em Abstract:} {\small For a symmetric kernel $k:X\times X
\rightarrow \mathbb{R}\cup\{+\infty\}$ on a locally compact
Hausdorff space $X$, we investigate the asymptotic behavior of
greedy $k$-energy points $\{a_{i}\}_{1}^{\infty}$ for a compact
subset $A\subset X$ that are defined inductively by selecting
$a_{1}\in A$ arbitrarily and $a_{n+1}$ so that
$\sum_{i=1}^{n}k(a_{n+1},a_{i})=\inf_{x\in
A}\sum_{i=1}^{n}k(x,a_{i})$. We give sufficient conditions under
which these points (also known as Leja points) are asymptotically
energy minimizing (i.e. have energy $\sum_{i\neq
j}^{N}k(a_{i},a_{j})$ as $N\rightarrow\infty$ that is
asymptotically the same as $\mathcal{E}(A,N):=\min\{\sum_{i\neq
j}k(x_{i},x_{j}):x_{1},\ldots,x_{N}\in A\}$), and have asymptotic
distribution equal to the equilibrium measure for $A$. For the
case of Riesz kernels $k_{s}(x,y):=|x-y|^{-s}$, $s>0$, we show
that if $A$ is a rectifiable Jordan arc or closed curve in
$\mathbb{R}^{p}$ and $s>1$, then greedy $k_{s}$-energy points are
\textit{not} asymptotically energy minimizing, in contrast to the
case $s<1$. (In fact we show that \textit{no} sequence of points
can be asymptotically energy minimizing for $s>1$.) Additional
results are obtained for greedy $k_{s}$-energy points on a sphere,
for greedy best-packing points, and for weighted Riesz kernels.}
\\



\section{{\large Introduction, background results and notation}}
The aim of this paper is to study asymptotic properties of special
types of extremal point configurations which we shall call
\textit{greedy energy points}. As the name suggests, these
configurations are generated by a greedy algorithm which is, in
fact, an energy minimizing construction. The notion of energy that
we refer to will be specified shortly. We focus on two aspects:
the asymptotic behavior of their energy and their limiting
distributions, as their cardinality approaches infinity. In many
aspects they are similar to minimal (non-greedy) energy
configurations, which are those with smallest possible energy. But
we will also show that in some situations the behavior of greedy
points differs significantly from that of minimal energy points.

Part of the results in this paper are presented in the abstract
setting of locally compact Hausdorff (LCH) spaces. Potential
theory on LCH spaces was developed by Choquet \cite{Choquet2},
\cite{Choquet}, Fuglede \cite{Fuglede} and Ohtsuka \cite{Ohtsuka}.
Recently Zorii \cite{Zorii}, \cite{Zorii2} has studied properties
of potentials with external fields in this context.

We also investigate greedy configurations in $\mathbb{R}^{p}$,
interacting through the so-called Riesz potential $V=1/r^{s}$,
where $s>0$ and $r$ denotes Euclidean distance, as well as greedy
`best-packing' points that are chosen to maximize the minimum
distance to previously selected points.

We next introduce the basic notions necessary to describe our
results. We will also present in this section some background
material.

Let $X$ denote a LCH space containing infinitely many points. A
\textit{kernel} in $X$ is, by definition, a lower semicontinuous
function (l.s.c.) $k:X \times X \rightarrow
\mathbb{R}\cup\{+\infty\}$. It is called \textit{positive} if
$k(x,y)\geq 0$ for all $x, y\in X$.

Given a set $\omega_N=\{x_1,\ldots,x_N\}$ of $N$ ($N\geq 2$)
points in $X$, not necessarily distinct, the \textit{discrete
energy} of $\omega_N$ is defined by
\[E(\omega_N):=\sum_{1\leq i\neq j\leq N}k(x_{i},x_{j})
=\sum_{i=1}^{N}\sum_{j=1,j\neq i}^{N}k(x_{i},x_{j})\,.\]

If the kernel is \textit{symmetric}, i.e., $k(x,y)=k(y,x)$ for all
$x, y\in X$, we may also write
\[E(\omega_N)=2 \sum_{1\leq i < j\leq N}k(x_{i},x_{j})\,.\]
An important notational convention that we will use throughout
this paper is the following: if $F\subset X$ is a set indexed by
some index set $I$, the expression $\card(F)$ will represent the
cardinality of $I$.

For a set $A\subset X$, the \textit{$N$-point energy} of $A$ is
given by
\begin{equation}\label{defnminimalenerg}
\mathcal{E}(A,N):=\inf\{E(\omega_{N}):\omega_N \subset A,
\,\,\card(\omega_N)=N\}\,.
\end{equation}
We say that $\omega_{N}^{*}\subset A$ is an \textit{optimal
$N$-point configuration} on $A$ if
\[E(\omega_{N}^{*})=\mathcal{E}(A,N)\,.\]
When $A$ is compact, such a configuration always exists by the
lower semicontinuity of $k$. In order to study the asymptotic
behavior of the sequence $E(\omega_{N}^{*})$ we need to introduce
the continuous counterparts of the above notions.

Let $\mathcal{M}(A)$ denote the linear space of all real-valued
Radon measures that are compactly supported on $A$, and let
$\mathcal{M}^{+}(A):=\{\mu\in\mathcal{M}(A): \mu\geq 0\}$. We also
introduce the class
$\mathcal{M}_{1}(A):=\{\mu\in\mathcal{M}^{+}(A): \mu(X)=1\}$.
Given a measure $\mu\in\mathcal{M}(A)$, the \textit{continuous
energy} of $\mu$ is the double integral
\begin{equation}\label{definicionenergia}
W(\mu):=\int\int k(x,y)\,d\mu(x)\,d\mu(y)\,.
\end{equation}
The function
\begin{equation}\label{defnpotential}
U^{\mu}(x):=\int k(x,y) \,d\mu(y)
\end{equation}
is called the \textit{potential} of $\mu$. Since any l.s.c.
function is bounded below on compact sets, the above integrals are
well-defined, although they may attain the value $+\infty$.

We say that $k$ satisfies the \textit{maximum principle} if for
every measure $\mu\in\mathcal{M}_{1}(A)$,
\begin{equation}\label{maximumprinciple}
\sup_{x\in \supp(\mu)}U^{\mu}(x)=\sup_{x\in X}U^{\mu}(x)\,.
\end{equation}

The quantity $w(A):=\inf\{W(\mu):\mu\in\mathcal{M}_{1}(A)\}$ plays
an important role in potential theory and is called the
\textit{Wiener energy} of $A$. The \textit{capacity} of $A$ is
defined as $\rm{cap}$$_{k}(A):=w(A)^{-1}$ if $k$ is positive, and
otherwise, it is defined as $\rm{cap}$$_{k}(A):=\exp(-w(A))$. A
property is said to hold \textit{quasi-everywhere} (q.e.), if the
exceptional set has Wiener energy $+\infty$.

Given a net $\{\mu_{\alpha}\}\subset\mathcal{M}(A)$, we say that
$\{\mu_{\alpha}\}$ converges in the \textit{weak-star topology} to
a measure $\mu\in\mathcal{M}(A)$ when
\[\lim_{\alpha}\int f \,d\mu_{\alpha}=\int f \,d\mu\,,
\qquad \mbox{for all}\quad f\in C_{c}(A)\,,\] where $C_{c}(A)$
denotes the space of compactly supported continuous functions on
$A$. We will use the notation
\[
\mu_{\alpha}\stackrel{*}{\longrightarrow}\mu
\]
to denote the weak-star convergence of measures. If $A$ is
compact, we know by the Banach-Alaoglu theorem that
$\mathcal{M}_{1}(A)$ equipped with the weak-star topology is
compact.

If $w(A)<\infty$, a measure $\mu\in\mathcal{M}_{1}(A)$ satisfying
the property $W(\mu)=w(A)$ is called an \textit{equilibrium
measure}. The existence of such a measure is guaranteed by the
lower semicontinuity of $k$ and the compactness of
$\mathcal{M}_{1}(A)$ (see Theorem 2.3 in \cite{Fuglede}). However,
uniqueness does not always hold.

The following result is due to G. Choquet \cite{Choquet}, and it
is central in this theory.
\begin{theo}\label{Choquettheo}
Let $k$ be an arbitrary kernel and $A\subset X$ be a compact set.
If $\{\omega_{N}^{*}\}$ is a sequence of optimal $N$-point
configurations on $A$, then
\begin{equation}\label{Choquetasymp}
\lim_{N\rightarrow\infty}\frac{E(\omega_{N}^{*})}{N^{2}}=w(A)\,.
\end{equation}
\end{theo}
The following variation of Theorem \ref{Choquettheo} was obtained
by Farkas and Nagy \cite{FarkasNagy}.
\begin{theo}\label{FNtheo}
Assume that the kernel $k$ is positive and is finite on the
diagonal, i.e., $k(x,x)<+\infty$ for all $x\in X$. Then for
arbitrary sets $A\subset X$,
\[
\lim_{N\rightarrow\infty}\frac{\mathcal{E}(A,N)}{N^{2}}=w(A)\,,
\]
where $\mathcal{E}(A,N)$ is defined by
$($$\ref{defnminimalenerg}$$)$.
\end{theo}

In this paper we study an alternative construction of points
obtained by means of a ``greedy'' algorithm.

\begin{defn}\label{greedypoints}
Let $k:X \times X \rightarrow \mathbb{R}\cup\{+\infty\}$ be a
symmetric kernel on a LCH space $X$, and let $A\subset X$ be a
compact set. A sequence $(a_{n})_{n=1}^{\infty}\subset A$ is
called a \textit{greedy $k$-energy sequence} on $A$ if it is
generated in the following way:
\begin{itemize}
\item $a_{1}$ is selected arbitrarily on $A$. \item Assuming that
$a_{1},\ldots,a_{n}$ have been selected, $a_{n+1}$ is chosen to
satisfy
\begin{equation}\label{eqnuevo4}
\sum_{i=1}^{n}k(a_{n+1},a_{i})= \inf_{x\in
A}\sum_{i=1}^{n}k(x,a_{i})\,,
\end{equation}
for every $n\geq 1$.
\end{itemize}

We remark that the choice of $a_{n+1}$ is not unique in general.
We will use the notation
\[\alpha_{N,k}:=\{a_{1},\ldots,a_{N}\}\]
to denote the set of the first $N$ points of this sequence. It is
significantly easier to obtain numerically these configurations
rather than optimal $N$-point configurations, since in order to
obtain the former we have to minimize a functional of one variable
instead of $N$ variables.
\end{defn}

It was shown by Fuglede (see Theorem 2.4 in \cite{Fuglede}) that
if $k$ is symmetric and $A\subset X$ is compact, every
$\mu\in\mathcal{M}_{1}(A)$ that has minimal energy satisfies the
inequality $U^{\mu}(x)\leq w(A)$ for all $x\in\supp(\mu)$. The
\textit{essential support} of $\mu$ is the set
\begin{equation}\label{defnessential}
S_{\mu}^{*}:=\{x\in A: U^{\mu}(x)\leq w(A)\}\,.
\end{equation}
Hence $\supp(\mu)\subset S_{\mu}^{*}$.

The following is a restricted version of Definition
\ref{greedypoints}.
\begin{defn}\label{greedypoints2}
Under the same assumptions as Definition \ref{greedypoints},
assume that $w(A)<\infty$, and let $\mu\in\mathcal{M}_{1}(A)$ be
an equilibrium measure. A sequence
$(a_{n}=a_{n,k,\mu})_{n=1}^{\infty}\subset A$ is called a
\textit{greedy $(k,\mu)$-energy sequence} on $A$ if it is
generated in the following way:
\begin{itemize}
\item $a_{1}$ is selected arbitrarily on $S_{\mu}^{*}$.

\item Assuming that $a_{1},\ldots,a_{n}$ have been selected,
$a_{n+1}$ is chosen to satisfy $a_{n+1}\in S_{\mu}^{*}$ and
\[
\sum_{i=1}^{n}k(a_{n+1},a_{i})=\inf_{x\in
S_{\mu}^{*}}\sum_{i=1}^{n}k(x,a_{i})\] for every $n\geq 1$.
\end{itemize}
The set of the first $N$ points of this sequence is denoted by
$\alpha_{N,k,\mu}$.
\end{defn}

Albert Edrei \cite{Edrei} was probably the first person who
studied the point configurations $\alpha_{N,k}$ in the particular
case $X=\mathbb{C}$ and $k(x,y)=-\log(|x-y|)$. However, in the
literature these configurations are often called \textit{Leja
points}, in recognition of Leja's article \cite{Leja}. When the
kernel employed is the Green function or the Newtonian kernel
$k(x,y)=1/|x-y|$ in the unit sphere $S^{2}$, the configurations
$\alpha_{N,k}$ are also referred to as \textit{Leja-G\'{o}rski
points} (see \cite{Gotz} and references therein). In
\cite{Baglama}, certain configurations known as \textit{fast Leja
points} are introduced, and an algorithm is presented to compute
them. These configurations are defined over discretizations of
planar sets and the kernel employed is the logarithmic kernel. In
\cite{CD} a constrained energy problem for this kernel is
considered and associated \textit{constrained Leja points} are
introduced. We remark that Leja points are important in
interpolation theory because they provide a Newton-type
interpolation point scheme on the real line or complex plane.

A very relevant class of kernels is the so-called \textit{M. Riesz
kernels} in $X=\mathbb{R}^{p}$, which depend on a parameter $s$ in
$[0,+\infty)$. It is defined as follows:
\[k_{s}(x,y):=K(|x-y|;s)\,,\qquad x,y\in\mathbb{R}^{p}\,,\]
where $|\cdot|$ denotes the Euclidean norm and
\begin{equation}\label{defnRieszK}
K(t;s):=\left\{
\begin{array}{ccc}
t^{-s}, & \mbox{if} & s>0\,, \\
-\log(t), & \mbox{if} & s=0\,.
\end{array}
\right.
\end{equation}
We shall use the notations $I_{s}(\mu)$ and $U_{s}^{\mu}$ to
denote the energy (\ref{definicionenergia}) and potential
(\ref{defnpotential}) of a measure $\mu\in\mathcal{M}(A)$ with
respect to the Riesz $s$-kernel, and $w_{s}(A)$ to denote the
Wiener energy of a set $A$ in this new setting. We will also use
$E_{s}(\omega_{N})$ to represent the discrete energy of an
$N$-point configuration $\omega_{N}\subset\mathbb{R}^{p}$, and
\begin{equation}\label{defnminimalenergriesz}
\mathcal{E}_{s}(A,N):=\inf\{E_{s}(\omega_{N}):\omega_N \subset A,
\,\,\card(\omega_N)=N\}
\end{equation}
to denote the $N$-point Riesz $s$-energy of a compact set
$A\subset\mathbb{R}^{p}$. Additionally, greedy $k_{s}$-energy
configurations will be denoted by $\alpha_{N,s}$.

A few words about Riesz $s$-kernels are needed at this point. Let
$A\subset\mathbb{R}^{p}$ be compact, and $0\leq
s<$dim$_{\mathcal{H}}(A)$, where dim$_{\mathcal{H}}(A)$ denotes
the Hausdorff dimension of $A$ (which will be denoted by $d$
throughout the rest of this section). Then there is a unique
equilibrium measure $\lambda_{A,s}\in\mathcal{M}_{1}(A)$ with
finite energy, i.e., $I_{s}(\lambda_{A,s})=w_{s}(A)<+\infty$. On
the other hand, if $s\geq d$, then $I_{s}(\mu)=+\infty$ for all
$\mu\in\mathcal{M}_{1}(A)$. We refer the reader to Theorems 8.5
and 8.9 in \cite{Mattila} for justifications of these facts.

For $s<d$, Theorem \ref{Choquettheo} asserts that
\begin{equation}\label{energsmenosd}
\lim_{N\rightarrow\infty}\frac{E_{s}(\omega_{N,s}^{*})}{N^{2}}=I_{s}(\lambda_{A,s})\,,
\end{equation}
where $\{\omega_{N,s}^{*}\}$ denotes any sequence of optimal
$N$-point configurations on $A$ with respect to the Riesz
$s$-kernel. In addition (see \cite{Landkof}),
\[
\frac{1}{N}\sum_{x\in\omega_{N,s}^{*}}\delta_{x}\stackrel{*}{\longrightarrow}
\lambda_{A,s}\,,\qquad N\rightarrow\infty\,,
\]
where $\delta_{x}$ is the Dirac unit measure concentrated at $x$.
If $s\geq d$, then Theorem \ref{Choquettheo} tells us that
\[\lim_{N\rightarrow\infty}\frac{E_{s}(\omega_{N,s}^{*})}{N^{2}}=+\infty\,,\]
so the order of growth of $E_{s}(\omega_{N,s}^{*})$ is greater
than $N^{2}$.

Throughout the rest of the paper we denote by $\vol(B^{d})$ the
volume of the unit ball $B^{d}$ in $\mathbb{R}^{d}$, and
$\mathcal{H}_{d}$ represents $d$-dimensional Hausdorff measure in
$\mathbb{R}^{p}$ (normalized by the condition
$\mathcal{H}_{d}([0,1]^{d})=1$, where $[0,1]^{d}$ denotes here the
embedding of the $d$-dimensional unit cube in $\mathbb{R}^{p}$).
Regarding the case $s\geq d$, in \cite{HardinSaff} and \cite{BHS}
geometric measure theoretic tools were employed to obtain the
following result.

\begin{theo}\label{theoBHSunweighted}
Let A be a compact subset of a d-dimensional $C^{1}$-manifold in
$\mathbb{R}^{p}$. If $\{\omega_{N,d}^{*}\}$ is any sequence of
optimal $N$-point configurations on $A$ for $s=d$, then
\begin{equation}\label{energsiguald}
\lim_{N\rightarrow\infty}\frac{E_{d}(\omega_{N,d}^{*})}{N^{2}\log
N} =\frac{\vol(B^{d})}{\mathcal{H}_{d}(A)}\,.
\end{equation}
Furthermore, if $\mathcal{H}_{d}(A)>0$, any sequence
$\{\widetilde{\omega}_N\}$ of configurations on $A$ whose energies
satisfy $(\ref{energsiguald})$ is uniformly distributed with
respect to $\mathcal{H}_{d}$ in the sense that
\begin{equation}\label{eqdistunweighted}
\frac{1}{N}\sum_{x\in\,\widetilde{\omega}_N}\delta_{x}\stackrel{*}{\longrightarrow}
\frac{\mathcal{H}_{d}|_{A}}{\mathcal{H}_{d}(A)}\,, \qquad
N\rightarrow\infty\,.
\end{equation}

Assume now that $A\subset\mathbb{R}^{p}$ is a $d$-rectifiable
compact set, i.e., $A$ is the image of a bounded set in
$\mathbb{R}^{d}$ under a Lipschitz mapping. If
$\{\omega_{N,s}^{*}\}$ is any sequence of optimal $N$-point
configurations on $A$ for $s>d$, there holds
\begin{equation}\label{energsmayord}
\lim_{N\rightarrow\infty}\frac{E_{s}(\omega_{N,s}^{*})}{N^{1+s/d}}=
\frac{C_{s,d}}{\mathcal{H}_{d}(A)^{s/d}}\,,
\end{equation}
where $C_{s,d}>0$ is a constant independent of $A$ and $p$. In
addition, if $\mathcal{H}_{d}(A)>0$, any sequence of
configurations on $A$ whose energies satisfy
$(\ref{energsmayord})$ is uniformly distributed with respect to
$\mathcal{H}_{d}$.
\end{theo}

We remark that the constant $C_{s,d}$ equals $2\zeta(s)$ when
$d=1$, where $\zeta(s)$ is the classical Riemann zeta function, as
was proved in \cite{M-FMRS}.

\begin{defn}\label{AEM}
Let $A$ be a compact set of Hausdorff dimension $d$. A sequence of
point sets $\omega_{N}\subset A$, is said to be
\textit{asymptotically s-energy minimizing} on $A$
($\{\omega_{N}\}_{N}\in \,$AEM$(A;s)$) if it satisfies, with
$\omega_{N,s}^{*}$ replaced by $\omega_{N}$, the limit relation
(\ref{energsmenosd}), (\ref{energsiguald}) or
(\ref{energsmayord}), according to whether $s<d$, $s=d$, or $s>d$.
\end{defn}

In Section \ref{mainresults} we state and discuss our main
results. Their proofs are given in subsequent sections.

\section{{\large main results}}\label{mainresults}

\subsection{The Potential theoretic case: Sets of positive
capacity}\label{seccion1}

Let
\[
U_{n}(x):=\sum_{j=1}^{n-1}k(x,a_{j})\,, \qquad n\geq 2\,.
\]
Our first result on the asymptotic behavior of greedy sequences is
the following.

\begin{theo}\label{theogredymaximum}
Let $k:X \times X \rightarrow \mathbb{R}\cup\{+\infty\}$ be a
symmetric kernel on a $\rm{LCH}$ space $X$ that satisfies the
maximum principle. Assume $A\subset X$ is a compact set and
$\{\alpha_{N,k}\}$ is a greedy $k$-energy sequence on $A$. Then
\begin{itemize}
\item [(i)] the following limit holds:
\begin{equation}\label{asympmk2}
\lim_{N\rightarrow\infty}\frac{E(\alpha_{N,k})}{N^{2}}=w(A)\,;
\end{equation}
\item[(ii)] if $w(A)<\infty$ and the equilibrium measure
$\mu\in\mathcal{M}_{1}(A)$ is unique, it follows that
\begin{equation}\label{asympdistribmk2}
\frac{1}{N}\sum_{a\in\alpha_{N,k}}\delta_{a}\stackrel{*}
{\longrightarrow}\mu\,,\qquad N\rightarrow\infty\,;
\end{equation}
\item[(iii)] if $w(A)<\infty$, there holds
\begin{equation}\label{eqasympUn}
\lim_{n\rightarrow\infty}\frac{U_{n}(a_{n})}{n}=w(A)\,,
\end{equation}
where $a_{n}$ is the $n$-th element of the greedy $k$-energy
sequence.
\end{itemize}

Furthermore, if $w(A)<\infty$, the analogues of assertions $(i),
(ii)$, and $(iii)$ hold for any greedy $(k,\mu)$-energy sequence
on $A$ without assuming the maximum principle.
\end{theo}

Theorem \ref{theogredymaximum} generalizes a result due to Siciak
\cite{Siciak} (see Lemma 3.1) stated for Riesz potentials. For
sets of positive capacity, his result asserts that if
$A\subset\mathbb{R}^{p}$ is a compact set, $p-2\leq s<p$, $p\geq
2$, and $\{\alpha_{N,s}\}$ is a greedy $k_{s}$-energy sequence on
$A$, then (\ref{eqasympUn}) holds for $k=k_{s}$.

As a consequence of Theorem \ref{theogredymaximum}, we deduce the
following corollaries for Riesz kernels. Throughout this paper we
denote the $d$-dimensional unit sphere in $\mathbb{R}^{d+1}$ by
$S^{d}$.
\begin{co}\label{co3}
Let $d$ be a positive integer and $s\in[0,d)$. Then any greedy
$k_{s}$-energy sequence $\alpha_{N,s}\subset S^{d}$ is
$\AEM(S^{d};s)$ and the asymptotic
formula\,\footnotemark\footnotetext{We remark that for $d=1$ and
$s=0$ we have $\mathcal{E}_{0}(S^{1},N)=-N\log(N)$, $N\geq 2$,
(cf. \cite{BrHS}).}
\begin{equation}\label{asympgreedysld}
\lim_{N\rightarrow\infty}\frac{E_{s}(\alpha_{N,s})}{N^{2}}=\left\{
\begin{array}{ccc}
\frac{\Gamma((d+1)/2)\Gamma(d-s)}{\Gamma((d-s+1)/2)\Gamma(d-s/2)}, & \mbox{if} & 0<s<d\,, \\
\\
-\log(2)+\frac{1}{2}(\psi(d)-\psi(d/2)), & \mbox{if} & s=0\,,
\end{array}
\right.
\end{equation}
holds, where $\psi(x):=\Gamma'(x)/\Gamma(x)$ denotes the digamma
function. In addition,
\begin{equation}\label{asympdistriesz}
\frac{1}{N}\sum_{a\in\alpha_{N,s}}\delta_{a}
\stackrel{*}{\longrightarrow}\sigma_{d}\,,\qquad
N\rightarrow\infty\,,
\end{equation}
where $\sigma_{d}$ is the normalized Lebesgue measure on $S^{d}$.
\end{co}

\begin{co}\label{corint}
Let $\alpha_{N,s}$ be any greedy $k_{s}$-energy sequence on
$[-1,1]$. For $s\in[0,1)$, this sequence is \AEM$([-1,1];s)$,
which means that
\begin{equation}\label{asympenergint}
\lim_{N\rightarrow\infty}\frac{E_{s}(\alpha_{N,s})}{N^{2}}=\left\{
\begin{array}{ccc}
\frac{\sqrt{\pi}\,\Gamma(1+s/2)}{\cos(\pi s/2)\Gamma((1+s)/2)}, & \mbox{if} & 0<s<1\,, \\
\\
\log(2), & \mbox{if} & s=0\,.
\end{array}
\right.
\end{equation}
Furthermore,
\[\frac{1}{N}\sum_{a\in\alpha_{N,s}}\delta_{a}\stackrel{*}{\longrightarrow}
\frac{c_{s}}{(1-x^{2})^{(1-s)/2}}\,dx\,,\qquad x\in[-1,1]\,,\quad
N\rightarrow\infty\,,\] where $c_{s}$ is a normalizing constant.
\end{co}

Our next result concerns second-order asymptotics for Riesz energy
on the unit circle. It is known that if $s\in(0,1)$, then the
following limit holds (see \cite{BrHS}).
\begin{equation}\label{nextordtermminimal}
\lim_{N\rightarrow\infty}\frac{\mathcal{E}_{s}(S^{1},N)-I_{s}(\sigma)N^{2}}
{N^{1+s}}=\frac{2\zeta(s)}{(2\pi)^{s}}\,,
\end{equation}
where $\mathcal{E}_{s}(S^{1},N)$ denotes (see
(\ref{defnminimalenergriesz})) the $N$-point minimal Riesz
$s$-energy of $S^{1}$, and $\zeta(s)$ is the analytic extension of
the classical Riemann zeta function. We know by Corollary
\ref{co3} that all greedy $k_{s}$-energy sequences are
$\AEM(S^{1};s)$ when $s\in(0,1)$. Nevertheless, the expression
(\ref{nextorderasymp}) below shows that in terms of second-order
asymptotics greedy $k_{s}$-energy sequences and optimal $N$-point
configurations for $s\in(0,1)$ behave differently.
\begin{prop}\label{theonoa1}
Let $s\in(0,1)$ and consider an arbitrary greedy $k_{s}$-energy
sequence $\{\alpha_{N,s}\}_{N}$ on $S^{1}$. Then the following
next order asymptotics holds:
\begin{equation}\label{nextorderasymp}
\lim_{n\rightarrow\infty}\frac{E_{s}(\alpha_{3\cdot
2^{n},s})-I_{s}(\sigma)(3\cdot 2^{n})^{2}}{(3\cdot 2^{n})^{1+s}}
=f(s)\frac{2\zeta(s)}{(2\pi)^{s}}\,,
\end{equation}
where $f(s)=\frac{1}{2}(\frac{4}{3})^{1+s}+(\frac{1}{3})^{1+s}<1$
for $s\in(0,1)$, $\zeta(s)$ is the analytic extension of the
classical Riemann zeta function, and $\sigma$ is the normalized
arclength measure on $S^{1}$.
\end{prop}

If $s\in(0,1)$, then $\zeta(s)<0$, and therefore
$f(s)\frac{2\zeta(s)}{(2\pi)^{s}}>\frac{2\zeta(s)}{(2\pi)^{s}}$.
Hence we obtain the following

\begin{co}\label{conseqsoa}
For all $s\in(0,1)$ and for any greedy $k_{s}$-energy sequence
$\{\alpha_{N,s}\}_{N}$ on $S^{1}$, the sequence
\[
\frac{E_{s}(\alpha_{N,s})-I_{s}(\sigma)N^{2}}{N^{1+s}}
\]
is not convergent.
\end{co}

\vspace{0.1cm}

\noindent\textbf{Remark:} It is well-known that on $S^{1}$ the
minimal $N$-point Riesz $s$-energy $\mathcal{E}_{s}(S^{1},N)$ is
attained only by configurations consisting of $N$ equally spaced
points, and this property holds for every $s\geq 0$. We will show
(see Lemma \ref{lem4}) that for such $s$ greedy configurations
$\alpha_{2^{n},s}$ on $S^{1}$ are formed by $2^{n}$ equally spaced
points.

\subsection{The Hypersingular Case: Sets of Capacity Zero}\label{subseccionprin}

\subsubsection{Greedy $k_s$-energy sequences on $S^{1}$}\label{subseccion2}

In this subsection we present some results about the asymptotic
behavior of $E_{s}(\alpha_{N,s})$ for greedy $k_{s}$-energy
sequences on $S^{1}$ when $s\geq1$. As we shall see in Theorem
\ref{theosm1}, greedy $k_s$-energy sequences on $S^{1}$ are not
AEM$(S^{1};s)$ for $s>1$, which is perhaps a surprising result. We
conclude that the behavior of $E_{s}(\alpha_{N,s})$ exhibits a
transition at $s=1$, the Hausdorff dimension of $S^{1}$, since as
we saw in the previous section greedy $k_{s}$-energy sequences are
AEM$(S^{1};s)$ for $s<1$.

\vspace{0.2cm}

\noindent\textbf{Remark:} It follows from the geometric lemmas
proved in Section \ref{prueba1} that greedy $k_{s}$-energy
sequences $\alpha_{N,s}$ on $S^{1}$ are independent of $s$, i.e.,
once the points $a_{1},\ldots,a_{n}$ have been selected, the
choice of $a_{n+1}$ is independent of the value of $s$ and depends
only on the position of the first $n$ points of the sequence. As a
consequence we will denote greedy $k_{s}$-energy sequences on
$S^{1}$ by $\alpha_{N}$ instead of $\alpha_{N,s}$.

\vspace{0.2cm}

In \cite{M-FMRS} (see Theorem 3.1) it was proved that if $\Gamma$
is a rectifiable Jordan arc, then for $s>1$,
\begin{equation}\label{eqM-FMRS}
\lim_{N\rightarrow\infty}\frac{E_{s}(\omega_{N,s}^{*})}{N^{1+s}}=
\frac{2\zeta(s)}{\mathcal{H}_{1}(\Gamma)^{s}}\,,
\end{equation}
and if $s=1$,
\begin{equation}\label{eqM-FMRS2}
\lim_{N\rightarrow\infty}\frac{E_{1}(\omega_{N,1}^{*})}{N^{2}\log
N}= \frac{2}{\mathcal{H}_{1}(\Gamma)}\,,
\end{equation}
where $\{\omega_{N,s}^{*}\}_{N}$ is any sequence of optimal
$N$-point configurations with respect to the Riesz $s$-kernel.

We remind the reader that by $\mathcal{E}_{s}(S^{1},N)$ we denote
the $N$-point Riesz $s$-energy of $S^{1}$ (see
(\ref{defnminimalenergriesz})). As it was observed previously,
optimal $N$-point configurations on $S^{1}$ consist precisely of
$N$ equally spaced points, and this property holds for all values
of $s\in[0,\infty)$. From (\ref{eqM-FMRS}) we have
\begin{equation}\label{asympcircle}
\lim_{N\rightarrow\infty}\frac{\mathcal{E}_{s}(S^{1},N)}{N^{1+s}}=
\frac{2\zeta(s)}{(2\pi)^{s}}\,.
\end{equation}

By Corollary \ref{co3} and Theorem \ref{theoweight} (see
Subsection \ref{subseccion3}) we know that if $s\in[0,d]$, then
any greedy $k_{s}$-energy sequence $\{\alpha_{N,s}\}$ on $S^{d}$
is AEM$(S^{d};s)$. However the situation changes when $s>1$ on
$S^1$.
\begin{prop}\label{theosm1}
For $s>1$, any greedy $k_{s}$-energy sequence
$\{\alpha_{N,s}\}_{N}$ on $S^{1}$ is not asymptotically $s$-energy
minimizing. In fact, the subsequence $\alpha_{3\cdot2^{n},s}$
satisfies
\[\lim_{n\rightarrow\infty}\frac{E_{s}(\alpha_{3\cdot2^{n},s})}{(3\cdot2^{n})^{1+s}}=f(s)\frac{2\zeta(s)}{(2\pi)^{s}}\,,\]
where $f(s)=\frac{1}{2}(\frac{4}{3})^{1+s}+(\frac{1}{3})^{1+s}>1$
for all $s>1$.
\end{prop}

As in the previous section, we want to describe the difference in
terms of second-order asymptotics between greedy $k_{s}$-energy
sequences and optimal $N$-point configurations when $s=1$. The
following formula holds (see \cite{BrHS}):
\begin{equation}\label{eqnextordminseq1}
\lim_{N\rightarrow\infty}\frac{\mathcal{E}_{1}(S^{1},N)-\frac{1}{\pi}N^{2}\log
N} {N^{2}}=\frac{1}{\pi}(\gamma-\log(\pi/2))\,,
\end{equation}
where
$\gamma=\lim_{M\rightarrow\infty}(1+\frac{1}{2}+\cdots+\frac{1}{M}-\log
M)$ denotes the Euler-Mascheroni constant.
\begin{prop}\label{theonoa2}
For any greedy $k_{1}$-energy sequence $\{\alpha_{N}\}_{N}$ on
$S^{1}$ we have
\begin{equation}\label{eqnextordgreedseq1}
\lim_{n\rightarrow\infty}\frac{E_{1}(\alpha_{3\cdot
2^{n}})-\frac{1}{\pi}(3\cdot 2^{n})^{2}\log(3\cdot 2^{n})}
{(3\cdot
2^{n})^{2}}=\frac{1}{\pi}(\gamma-\log(\pi/2)+\log(2^{\frac{16}{9}}/3))\,.
\end{equation}
\end{prop}

\begin{co}\label{conuevo}
For any greedy $k_{1}$-energy sequence $\{\alpha_{N}\}_{N}$ on
$S^{1}$, the sequence
\[
\frac{E_{1}(\alpha_{N})-\frac{1}{\pi}N^{2}\log N} {N^{2}}
\]
is not convergent.
\end{co}

\subsubsection{$k_s$-Energy of sequences on Jordan arcs or curves
in $\mathbb{R}^{p}$ for $s\geq 1$ and
best-packing}\label{subseccion4} Throughout this subsection, by a
Jordan arc in $\mathbb{R}^{p}$ we understand a set homeomorphic to
a closed segment. A closed Jordan curve refers to a set
homeomorphic to a circle.

Our main result states that for $s>1$ it is \textit{not} possible
to find \textit{any} sequence of points on a Jordan arc or curve
that is asymptotically $s$-energy minimizing.

\begin{theo}\label{theoconjforarcs}
Let $\{x_{k}\}_{k=0}^{\infty}\subset\Gamma$ be an arbitrary
sequence of distinct points, where $\Gamma$ is a rectifiable
Jordan arc or closed Jordan curve in $\mathbb{R}^{p}$. Set
$\mathcal{X}_{n}:=\{x_{k}\}_{k=0}^{n}$. Then
$\{\mathcal{X}_{n}\}_{n}\notin \AEM(\Gamma;s)$ for all $s>1$. In
particular, $\{\alpha_{N,s}\}\notin\AEM(\Gamma;s)$ for any greedy
$k_{s}$-energy sequence on $\Gamma$ when $s>1$.
\end{theo}

The next result shows that, in contrast to the case $s>1$, for
$s=1$ greedy $k_{1}$-energy sequences on $S^{1}$ are
$\AEM(S^{1};1)$. More generally, we shall prove this fact for
\textit{smooth} Jordan arcs or curves $\Gamma$ by which we mean
that the natural parametrization
$\Phi:[0,L]\longrightarrow\Gamma$, where
$L=\mathcal{H}_{1}(\Gamma)$, is of class $C^{1}$ and $\Phi'(t)\neq
\textbf{0}$ for all $t\in[0,L]$.
\begin{theo}\label{theo2}
Let $\Gamma\subset\mathbb{R}^{p}$ be a smooth Jordan arc or closed
curve, and let $s=d=1$. Then any greedy $k_{1}$-energy sequence
$\{\alpha_{N,1}\}$ on $\Gamma$ is \AEM$(\Gamma;1)$, i.e.
\begin{equation}\label{eq:asymparc}
\lim_{N\rightarrow\infty}\frac{E_{1}(\alpha_{N,1})}{N^{2}\log N}
=\frac{2}{\mathcal{H}_{1}(\Gamma)}\,.
\end{equation}
Furthermore,
\begin{equation}\label{eq:distasymparc}
\frac{1}{N}\sum_{a\in\alpha_{N,1}}\delta_{a}\stackrel{*}
{\longrightarrow}\frac{\mathcal{H}_{1}|_{\Gamma}}
{\mathcal{H}_{1}(\Gamma)}\,,\qquad N\rightarrow\infty\,.
\end{equation}
\end{theo}

For the analogous result for greedy $k_{d}$-energy on the unit
sphere $S^{d}\subset\mathbb{R}^{d+1}$, see Theorem
\ref{theoweight}.

We next consider best-packing configurations. For a collection of
$N$ distinct points
$\omega_{N}=\{x_{1},\ldots,x_{N}\}\subset\mathbb{R}^{p}$ we set
\[
\delta(\omega_{N}):=\min_{1\leq i\neq j\leq N}|x_{i}-x_{j}|\,,
\]
and for an infinite set $A\subset\mathbb{R}^{p}$, we let
\[
\delta_{N}(A):=\sup\{\delta(\omega_{N}):\,\omega_{N}\subset A,\,
\card(\omega_{N})=N\}
\]
be the \textit{best-packing distance} of $N$-point configurations
on $A$. In \cite{BHS2} it is shown (see Theorem 2.2) that if
$A=\Gamma$ is a rectifiable Jordan curve or arc in
$\mathbb{R}^{p}$,
\[
\lim_{N\rightarrow\infty}N\delta_{N}(\Gamma)=\mathcal{H}_{1}(\Gamma)\,.
\]

This fact leads us to the following.
\begin{defn}
Let $\Gamma\subset\mathbb{R}^{p}$ be a Jordan arc or curve, and
let $\omega_{N}\subset\Gamma$ be a sequence of $N$-point
configurations. We say that $\{\omega_{N}\}\in\AEM(\Gamma,\infty)$
if
\[
\lim_{N\rightarrow\infty}N\delta(\omega_{N})=\mathcal{H}_{1}(\Gamma)\,.
\]
\end{defn}
\begin{theo}\label{bestpackrevisited}
Let $\Gamma\subset\mathbb{R}^{p}$ be a rectifiable Jordan arc or
curve with length $L=\mathcal{H}_{1}(\Gamma)$, and let
$\{x_{k}\}_{k=0}^{\infty}\subset\Gamma$ be an arbitrary infinite
sequence such that $x_{i}\neq x_{j}$ if $i\neq j$. Set
$\mathcal{X}_{n}:=\{x_{0},\ldots,x_{n}\}$. Then
$\{\mathcal{X}_{n}\}\notin\AEM(\Gamma,\infty)$. In fact,
\begin{equation}\label{eqbestpack}
\liminf_{n\rightarrow\infty}n\,\delta(\mathcal{X}_{n})
\leq\frac{4+3\sqrt{2}}{4+4\sqrt{2}}\,L<L\,.
\end{equation}
Moreover, if
$c:=\limsup_{n\rightarrow\infty}n\,\delta(\mathcal{X}_{n})>\frac{2+\sqrt{2}}{4}L$,
then
\begin{equation}\label{eqbestpack2}
\liminf_{n\rightarrow\infty}n\,\delta(\mathcal{X}_{n})\leq\frac{L}{2}+\sqrt{c\,(L-c)}<c\,.
\end{equation}
In particular, if
$\limsup_{n\rightarrow\infty}n\,\delta(\mathcal{X}_{n})=L$, then
$\liminf_{n\rightarrow\infty}n\,\delta(\mathcal{X}_{n})\leq L/2$.
\end{theo}

In analogy with finite $s$, we define \textit{greedy best-packing
configurations} on an infinite compact set
$A\subset\mathbb{R}^{p}$ by selecting $a_{0}\in A$ and choosing
$a_{n}\in A$ so that
\[
\min_{0\leq i\leq n-1}|a_{n}-a_{i}|=\max_{x\in A}\,\,\min_{0\leq
i\leq n-1}|x-a_{i}|\,.
\]
Such points are referred to in \cite{DeMarchi} as
\textit{Leja-Bos} points. Theorem \ref{bestpackrevisited} shows
that such points are not asymptotically optimal on rectifiable
Jordan arcs or curves.

In \cite{DeMarchi} there appears a conjecture attributed to L. Bos
stating that if $A$ is a compact domain of $\mathbb{C}$, every
Leja-Bos sequence $\{a_{n}\}_{n=0}^{\infty}$ on $A$ with
$|a_{0}|=\max\{|x|:x\in A\}$ is asymptotically uniformly
distributed. We wish to point out that this conjecture is false as
the following result asserts (see also Figure 1 in Section
\ref{seccion5}).

\begin{prop}\label{counterexample}
There exist greedy best-packing sequences on $[0,1]$ and $[0,1]^2$
that are not asymptotically uniformly distributed.
\end{prop}

It is obvious, however, that greedy best-packing sequences are
\textit{dense} in $A$.

\subsubsection{Weighted Riesz potentials}\label{subseccion3} In
this subsection we will consider the notion of weighted discrete
Riesz energy introduced in \cite{BHS}. We reproduce here the main
definitions.

\begin{defn}
Let $A\subset \mathbb{R}^{p}$ be an infinite compact set whose
$d$-dimensional Hausdorff measure $\mathcal{H}_{d}(A)$ is finite.
A symmetric function $w:A \times A \longrightarrow [0,\infty]$ is
called a CPD-\textit{weight function} on $A\times A$ if
\begin{itemize}
\item $w$ is continuous (as a function on $A\times A$) at
$\mathcal{H}_{d}$-almost every point of the diagonal
$D(A):=\{(x,x):x\in A\}$,

\item there is some neighborhood $G$ of $D(A)$ (relative to
$A\times A$) such that $\inf_{G}w>0$, and

\item $w$ is bounded on any closed subset $B\subset A\times A$
such that $B\cap D(A)=\emptyset$.
\end{itemize}
\end{defn}
The term CPD stands for (almost) continuous and positive on the
diagonal.
\begin{defn}
Let $s>0$. Given a collection of $N$ ($N\geq 2$) points
$\omega_N:=\{x_1,\ldots,x_N\}\subset A$, the \textit{weighted
Riesz $s$-energy of $\omega_N$} is defined by
\[E_{s}^{w}(\omega_N):=\sum_{1\leq i\neq j\leq N}
\frac{w(x_{i},x_{j})}{|x_{i}-x_{j}|^{s}}\,,\] while the
\textit{$N$-point weighted Riesz $s$-energy of $A$} is given by
\[\mathcal{E}_{s}^{w}(A,N):=\inf\{E_{s}^{w}(\omega_N):\omega_N\subset A,
\,\,\card(\omega_N)=N\}\,.\] The \textit{weighted Hausdorff
measure $\mathcal{H}_{d}^{s,w}$} on Borel sets $B\subset A$ is
defined by
\[\mathcal{H}_{d}^{s,w}(B):=\int_{B}(w(x,x))^{-d/s}d\mathcal{H}_{d}(x)\,.\]
\end{defn}

The following result about the asymptotic behavior of
$\{\mathcal{E}_{s}^{w}(A,N)\}_{N}$ was obtained in \cite{BHS}.
\begin{theo}\label{theoBHS}
Let $A$ be a compact subset of a $d$-dimensional $C^{1}$-manifold
in $\mathbb{R}^{p}$ and assume that $w:A\times A\rightarrow
[0,\infty]$ is a $\CPD$-weight function on $A\times A$. Then
\begin{equation}\label{theoBHSeq1}
\lim_{N\rightarrow\infty}\frac{\mathcal{E}_{d}^{w}(A,N)}{N^{2}\log
N} =\frac{\vol(B^{d})}{\mathcal{H}_{d}^{d,w}(A)}\,,
\end{equation}
Furthermore, if $\mathcal{H}_{d}(A)>0$ and
$\{\widetilde{\omega}_N\}$ is a sequence of configurations on $A$
satisfying $(\ref{theoBHSeq1})$, with $\mathcal{E}_{d}^{w}(A,N)$
replaced by $E_{d}^{w}(\widetilde{\omega}_N)$, then
\begin{equation}\label{theoBHSeq2}
\frac{1}{N}\sum_{x\in\,\widetilde{\omega}_N}^{N}\delta_{x}\stackrel{*}{\longrightarrow}
\frac{\mathcal{H}_{d}^{d,w}|_{A}}{\mathcal{H}_{d}^{d,w}(A)}
\,,\qquad N\rightarrow\infty\,.
\end{equation}

Assume now that $A\subset\mathbb{R}^{p}$ is a $d$-rectifiable set.
Then for $s>d$,
\begin{equation}\label{theoBHSeq3}
\lim_{N\rightarrow\infty}\frac{\mathcal{E}_{s}^{w}(A,N)}{N^{1+s/d}}=\frac{C_{s,d}}
{[\mathcal{H}_{d}^{s,w}(A)]^{s/d}}\,,
\end{equation}
where $C_{s,d}$ is the same positive constant that appears in
Theorem $\ref{theoBHSunweighted}$. In addition, if
$\mathcal{H}_{d}(A)>0$, any sequence $\{\widetilde{\omega}_N\}$ of
configurations on $A$ satisfying $(\ref{theoBHSeq3})$ with
$\mathcal{E}_{s}^{w}(A,N)$ replaced by
$E_{s}^{w}(\widetilde{\omega}_N)$ also satisfies
$(\ref{theoBHSeq2})$.
\end{theo}

\begin{defn}\label{defngwses}
Let $w$ be a lower semicontinuous CPD-weight function on $A\times
A$. A sequence $(a_{n})_{n=1}^{\infty}\subset A$ is called a
\textit{greedy $(w,s)$-energy sequence} on $A$ if it is generated
in the same way as generated in Definition \ref{greedypoints},
with $k(x,y):=w(x,y)/|x-y|^{s}$.
\end{defn}

The next result concerns greedy $(w,d)$-energy points on the unit
sphere $S^{d}\subset\mathbb{R}^{d+1}$.

\begin{theo}\label{theoweight}
Assume that $w:S^{d}\times S^{d}\rightarrow [0,\infty)$ is a
continuous function such that $w(x,x)>0$ for all $x\in S^{d}$. Let
$\{\alpha_{N,d}^{w}\}_{N}$ be an arbitrary greedy $(w,d)$-energy
sequence on $S^{d}$, $d\geq 1$. Then
\begin{equation}\label{eqtheoweight}
\lim_{N\rightarrow\infty}\frac{E_{d}^{w}(\alpha_{N,d}^{w})}
{N^{2}\log N}=\frac{\vol(B^{d})} {\mathcal{H}_{d}^{d,w}(S^{d})}\,,
\end{equation}
and therefore
\[\frac{1}{N}\sum_{a\in\alpha_{N,d}^{w}}\delta_{a}
\stackrel{*}{\longrightarrow}
\frac{\mathcal{H}_{d}^{d,w}|_{S^{d}}}
{\mathcal{H}_{d}^{d,w}(S^{d})}\,,\qquad N\rightarrow\infty\,.\] In
particular, any greedy $k_{d}$-energy sequence
$\{\alpha_{N,d}\}_{N}$ on $S^{d}$ is $\AEM(S^{d},d)$ and satisfies
$(\ref{asympdistriesz})$ for $s=d$.
\end{theo}

In the following result we consider greedy $(w,p)$-energy
sequences on sets in $\mathbb{R}^{p}$ with positive Lebesgue
measure.

\begin{theo}\label{theoweightset}
Let $A\subset\mathbb{R}^{p}$ be a compact set such that
$\mathcal{H}_{p}(A)>0$, and let $\{\alpha_{N,p}^{w}\}_{N}$ be an
arbitrary greedy $(w,p)$-energy sequence on $A$. Assume that
$w:A\times A\rightarrow [0,\infty)$ is a continuous function such
that $w(x,x)>0$ for all $x\in A$. Then
\begin{equation}\label{eqnuevo}
\lim_{N\rightarrow\infty}\frac{E_{p}^{w}(\alpha_{N,p}^{w})}
{N^{2}\log N}=\frac{\vol(B^{p})} {\mathcal{H}_{p}^{p,w}(A)}\,,
\end{equation}
and therefore
\begin{equation}\label{eqnuevo3}
\frac{1}{N}\sum_{a\in\alpha_{N,p}^{w}}\delta_{a}
\stackrel{*}{\longrightarrow}
\frac{\mathcal{H}_{p}^{p,w}|_{A}}{\mathcal{H}_{p}^{p,w}(A)}
\,,\qquad N\rightarrow\infty\,.
\end{equation}
In particular, any greedy $k_{p}$-energy sequence
$\{\alpha_{N,p}\}_{N}$ on $A$ is $\AEM(A;p)$ and is asymptotically
uniformly distributed with respect to $\mathcal{H}_{p}$.

\end{theo}

In view of Proposition \ref{theosm1}, it is not in general
possible to extend Theorem \ref{theoweight} to $s>d$. However, for
any compact set $A\subset\mathbb{R}^{p}$ with
$\mathcal{H}_{\delta}(A)>0$ (where $\delta>0$ is arbitrary, not
necessarily an integer), we can show that the order of growth of
$E_{s}^{w}(\alpha_{N,s}^{w})$ when $s>\delta$ ($s=\delta$) is at
most $N^{1+s/\delta}$ ($N^{2}\log N$). Let
\[
\mathcal{H}_{\delta}^{\infty}(A):=\inf\{\sum_{i}(\diam
\,G_{i})^{\delta}:A\subset \bigcup_{i}G_{i}\}\,,\qquad \delta>0\,.
\]
\begin{theo}\label{ordergrowth}
Let $0<\delta\leq p$. Assume that $A\subset\mathbb{R}^{p}$ is a
compact set such that $\mathcal{H}_{\delta}(A)>0$. Let $w$ be a
bounded lower semicontinuous $\CPD$-weight function on $A\times
A$. Consider an arbitrary greedy $(w,s)$-energy sequence
$\{\alpha_{N,s}^{w}\}_{N}\subset A$, for $s\geq \delta$. Then, for
$N\geq 2$
\[
E_{s}^{w}(\alpha_{N,s}^{w})\leq\left\{
\begin{array}{ccc}
M_{s,\delta,A}\,\|w\|\,\mathcal{H}_{\delta}^{\infty}(A)^{-s/\delta}N^{1+s/\delta},
& \mbox{if} & s>\delta\,,\\
\\
M_{\delta,A}\,\|w\|\,\mathcal{H}_{\delta}^{\infty}(A)^{-1}N^{2}\log
N, & \mbox{if} & s=\delta\,,
\end{array}
\right.
\]
where the constants $M_{s,\delta,A}>0$ and $M_{\delta,A}>0$ are
independent of $w$ and $N$, and $\|w\|:=\sup\{w(x,y):x,y\in A\}$.
\end{theo}

\begin{co}\label{coimp}
Let $A\subset\mathbb{R}^{p}$ be a $d$-rectifiable set. Suppose
$s>d$ and $w$ is a bounded lower semicontinuous $\CPD$-weight
function on $A\times A$. Consider an arbitrary greedy
$(w,s)$-energy sequence $\{\alpha_{N,s}^{w}\}_{N}\subset A$. Then
there are constants $C_{1}, C_{2}>0$ such that
\begin{equation}\label{eqprimera}
C_{1}\,N^{1+s/d}\leq E_{s}^{w}(\alpha_{N,s}^{w})\leq
C_{2}\,N^{1+s/d}\,.
\end{equation}
If $s=d$ and $A$ is assumed to be a compact subset of a
$d$-dimensional $C^{1}$-manifold, then there are constants $C_{3},
C_{4}>0$ such that
\begin{equation}\label{eqsegunda}
C_{3}\,N^{2}\log N\leq E_{d}^{w}(\alpha_{N,d}^{w})\leq
C_{4}\,N^{2}\log N\,,
\end{equation}
for any greedy $(w,d)$-energy sequence
$\{\alpha_{N,d}^{w}\}_{N}\subset A$.
\end{co}

\begin{co}\label{coimp2}
Let $A\subset\mathbb{R}^{p}$ be a $d$-rectifiable set. Suppose
$s>d$ and $w$ is a bounded lower semicontinuous $\CPD$-weight
function on $A\times A$. Consider an arbitrary greedy
$(w,s)$-energy sequence $\{a_{n}\}_{n=1}^{\infty}\subset A$. Then
$\{a_{n}\}_{n=1}^{\infty}$ is dense in $A$. If $s=d$ and $A$ is
assumed to be a compact subset of a $d$-dimensional
$C^{1}$-manifold, the same conclusion holds for any greedy
$(w,d)$-energy sequence. Taking $w\equiv 1$ the result is
applicable to greedy $k_{s}$-energy sequences.
\end{co}

We can slightly improve the density result in certain cases like a
real interval.

\begin{prop}\label{theodensint}
Let $[a,b]\subset\mathbb{R}$ and $s>1$. Assume that $w$ is a
bounded lower semicontinuous $\CPD$-weight function on
$[a,b]\times [a,b]$, and $(a_{n})_{n=1}^{\infty}$ is a greedy
$(w,s)$-energy sequence on $[a,b]$. If $I$ is any closed
subinterval of $[a,b]$, then
\begin{equation}\label{eq36}
\liminf_{N\rightarrow\infty}\frac{(\card\{1\leq n\leq N: a_{n}\in
I\})^{1+\frac{1}{s}}}{N}>0\,.
\end{equation}
\end{prop}

We finish this section remarking that some results about greedy
sequences in the context of external fields have been obtained by
the first author and will appear in a separate work.

\section{{\large Proofs of results from Section \ref{seccion1}}}\label{seccion3}

\noindent\textbf{Proof of Theorem \ref{theogredymaximum}.} Assume
first that $w(A)<\infty$, $\mu\in\mathcal{M}_{1}(A)$ is an
equilibrium measure, and $\{\alpha_{N,k,\mu}\}$ is an arbitrary
greedy $(k,\mu)$-energy sequence on $A$. If $a_{n}$ is the $n$-th
element of this sequence, it follows by definition that
\[
U_{n}(a_{n})\leq U_{n}(x)\,,\qquad\mbox{for all}\quad x\in
S_{\mu}^{*}\,,\quad n\geq 2\,.\] Hence, for any $x\in
S_{\mu}^{*}$,
\[
E(\alpha_{N,k,\mu})=\sum_{1\leq i\neq j\leq
N}k(a_{i},a_{j})=2\sum_{j=2}^{N}\sum_{i=1}^{j-1}k(a_{i},a_{j})
\]
\[
=2\sum_{j=2}^{N}U_{j}(a_{j})\leq
2\sum_{j=2}^{N}U_{j}(x)=2\sum_{j=2}^{N}\sum_{i=1}^{j-1}k(x,a_{i})\,.
\]
We now integrate the above inequality with respect to $\mu$ to
obtain
\[
E(\alpha_{N,k,\mu})\leq
2\sum_{j=2}^{N}\sum_{i=1}^{j-1}U^{\mu}(a_{i})\,.
\]
Taking into account that $U^{\mu}(a_{i})\leq w(A)$ for all $i$
($a_{i}\in S_{\mu}^{*}$) it follows that
\begin{equation}\label{eqmitad}
E(\alpha_{N,k,\mu})\leq N(N-1)\,w(A)\,.
\end{equation}
Now, if $\{\omega_{N}^{*}\}$ is a sequence of optimal $N$-point
configurations on $A$, then $E(\omega_{N}^{*})\leq
E(\alpha_{N,k,\mu})$ for all $N$. Therefore (\ref{asympmk2}) for
$\alpha_{N,k,\mu}$ is a consequence of (\ref{eqmitad}) and
(\ref{Choquetasymp}).

Consider the sequence of normalized counting measures
\[
\nu_{N}:=\frac{1}{N}\sum_{a\in\alpha_{N,k,\mu}}\delta_{a}\,,
\]
and assume that the equilibrium measure $\mu$ is unique. Let
$g_{n}:X\times X\rightarrow\mathbb{R}$ be a sequence of
non-decreasing continuous functions that converges pointwise to
$k$. We have
\[
\int\int g_{n}(x,y)\,d\nu_{N}(x)d\nu_{N}(y)=
\frac{1}{N^{2}}\sum_{i=1}^{N}\sum_{j=1}^{N}g_{n}(a_{i},a_{j})
\]
\[
=\frac{1}{N^{2}}\Big(\sum_{i=1}^{N}g_{n}(a_{i},a_{i})+\sum_{1\leq
i\neq j\leq N}g_{n}(a_{i},a_{j})\Big)
\]
\[
\leq\frac{1}{N^{2}}\Big(\sum_{i=1}^{N}g_{n}(a_{i},a_{i})+\sum_{1\leq
i\neq j\leq N}k(a_{i},a_{j})\Big)\,;
\]
hence
\begin{equation}\label{eq:rev}
\int\int g_{n}(x,y)d\nu_{N}(x)d\nu_{N}(y)\leq
\frac{\sum_{i=1}^{N}g_{n}(a_{i},a_{i})
+E(\alpha_{N,k,\mu})}{N^2}\,.
\end{equation} By the compactness of $A$ and the continuity of $g_{n}$, there
exists a constant $M_{n}>0$ such that
\[
\sum_{i=1}^{N}|g_{n}(a_{i},a_{i})|\leq N\,M_{n}\,.
\]
Therefore, for each fixed $n$ we have
\begin{equation}\label{eq:rev2}
\lim_{N\rightarrow\infty}\frac{\sum_{i=1}^{N}g_{n}(a_{i},a_{i})
+E(\alpha_{N,k,\mu})}{N^2}=w(A)\,.
\end{equation}
Let $\{\nu_{N}\}_{N\in\mathcal{N}}$ be a subsequence that
converges in the weak-star topology to a measure
$\lambda\in\mathcal{M}_{1}(A)$. Since $\nu_{N}\times\nu_{N}$
converges weak star to $\lambda\times\lambda$, we have
\[
\lim_{N\rightarrow\infty}\int\int
g_{n}(x,y)d\nu_{N}(x)d\nu_{N}(y)=\int\int
g_{n}(x,y)d\lambda(x)d\lambda(y)\,.
\]
Thus from (\ref{eq:rev}) and (\ref{eq:rev2}) we conclude that
\[\int\int g_{n}(x,y)d\lambda(x)d\lambda(y)\leq w(A)\,.\]
Now we let $n\rightarrow\infty$ to obtain
\[W(\lambda)=\int\int k(x,y)d\lambda(x)d\lambda(y)\leq w(A)\,.\]
It follows that $\lambda$ is an equilibrium measure. By hypothesis
there is only one equilibrium measure. Thus $\lambda=\mu$ and
(\ref{asympdistribmk2}) is proved for $\alpha_{N,k,\mu}$.

We next show (\ref{eqasympUn}) for $\alpha_{N,k,\mu}$. It is not
assumed now that the equilibrium measure is unique, and
$\alpha_{N,k,\mu}$ denotes a greedy $(k,\mu)$-energy sequence
associated with a certain equilibrium measure $\mu$. We know from
the first part of the proof that
\begin{equation}\label{eq:uno}
\lim_{N\rightarrow\infty}\frac{E(\alpha_{N,k,\mu})}{N^{2}}
=\lim_{N\rightarrow\infty}\frac{2\sum_{i=2}^{N}U_{i}(a_{i})}{N^{2}}=w(A)\,.
\end{equation}
For every $n\geq 1$,
\[
\frac{U_{n+1}(a_{n+1})}{n}=\inf_{x\in
S_{\mu}^{*}}\frac{1}{n}\sum_{i=1}^{n}k(x,a_{i})\,.
\]
Integrating this equality with respect to $\mu$ we get
\begin{equation}\label{eq:dos}
\frac{U_{n+1}(a_{n+1})}{n}\leq\frac{1}{n}\sum_{i=1}^{n}\int
k(x,a_{i}) \,d\mu(x)=\frac{1}{n}\sum_{i=1}^{n} U^{\mu}(a_{i})\leq
w(A)\,.
\end{equation}
On the other hand, for every $n\geq 2$,
\begin{equation}\label{eq:tres}
U_{n+1}(a_{n+1})\geq U_{n}(a_{n})+L\,,
\end{equation}
where $L:=\inf\{k(x,y):x,y\in S_{\mu}^{*}\}$. We may assume that
$L\leq -1$.

Let $\epsilon\in(0,1)$. Assume that $m$ is an integer such that
\begin{equation}\label{eq:cuatro}
\frac{U_{m+1}(a_{m+1})}{m}<w(A)-\epsilon\,.
\end{equation}
Applying (\ref{eq:tres}) repeatedly we obtain for
$(1+\epsilon/(3L))\,m\leq i\leq m$,
\[
\frac{U_{i+1}(a_{i+1})}{m}\leq w(A)-\epsilon-\frac{(m-i)L}{m}\leq
w(A)-\epsilon+\frac{\epsilon/3}{1+\epsilon/(3L)} \leq
w(A)-\frac{\epsilon}{2}\,,
\]
and so
\[
\frac{U_{i+1}(a_{i+1})}{i}\leq \frac{m}{i}(w(A)-\epsilon/2)\leq
\frac{m}{i}\,w(A)-\frac{\epsilon}{2}\,.
\]
Taking into account (\ref{eq:dos}) and the last inequality,
\begin{equation}\label{eq:siete}
\frac{2}{(m+1)\,m}\sum_{i=1}^{m}U_{i+1}(a_{i+1})\leq
\frac{2}{(m+1)\,m}\sum_{1\leq i<(1+\epsilon/(3L))m}i\,w(A)
\end{equation}
\[
+\frac{2}{(m+1)\,m}\sum_{(1+\epsilon/(3L))m\leq i\leq
m}m\,w(A)-\frac{\epsilon}{2}\frac{2}{(m+1)\,m}\sum_{(1+\epsilon/(3L))m\leq
i\leq m}i\,.
\]
Furthermore, it is easy to see that
\begin{equation}\label{eq:seis}
-\frac{\epsilon}{2}\frac{2}{(m+1)\,m}\sum_{(1+\epsilon/(3L))m\leq
i\leq m}i\leq
\frac{\epsilon^2}{6L(m+1)}\Big(1+2m+\frac{m\epsilon}{3L}\Big)
\end{equation}
\[
\leq \frac{\epsilon^{2}(1+\epsilon/(3L))}{6L}\,.
\]

If $w(A)\leq 0$, then
\[
\frac{2}{(m+1)\,m}\Big\{\sum_{1\leq
i<(1+\epsilon/(3L))m}i\,w(A)+\sum_{(1+\epsilon/(3L))m\leq i\leq
m}m\,w(A) \Big\}\leq w(A)
\]
and hence it follows from (\ref{eq:siete}) and (\ref{eq:seis})
that
\begin{equation}\label{eq:cinco}
\frac{2}{(m+1)\,m}\sum_{i=1}^{m}U_{i+1}(a_{i+1})\leq
w(A)+\frac{\epsilon^{2}(1+\epsilon/(3L))}{6L}\,.
\end{equation}
Since the right-hand side of (\ref{eq:cinco}) is a constant
strictly less than $w(A)$, by (\ref{eq:uno}) it follows that there
are only finitely many integers $m$ satisfying (\ref{eq:cuatro}).
This implies with (\ref{eq:dos}) that (\ref{eqasympUn}) holds.

Now assume that $w(A)>0$. It is easy to verify that
\[
\frac{2}{(m+1)\,m}\Big\{\sum_{1\leq
i<(1+\epsilon/(3L))m}i\,w(A)+\sum_{(1+\epsilon/(3L))m\leq i\leq
m}m\,w(A) \Big\}
\]
\[
\leq \Big(1+\frac{2}{m+1}+\frac{\epsilon}{3L(m+1)}
+\frac{\epsilon^{2}m}{9(m+1)L^{2}}\Big)w(A)\,,
\]
and so, from (\ref{eq:siete}) and (\ref{eq:seis}), we deduce that
\[
\frac{2}{(m+1)\,m}\sum_{i=1}^{m}U_{i+1}(a_{i+1})\leq
\Big(1+\frac{2}{m+1}+\frac{\epsilon}{3L(m+1)}+\frac{\epsilon^{2}m}{9(m+1)L^{2}}\Big)w(A)
\]
\[
+\frac{\epsilon^{2}(1+\epsilon/(3L))}{6L}\,.
\]
If we assume that there are infinitely many integers $m$
satisfying (\ref{eq:cuatro}), then applying the last inequality we
obtain
\begin{equation}\label{eq:ocho}
\limsup_{N\rightarrow\infty}\frac{2\sum_{i=2}^{N}U_{i}(a_{i})}{N^{2}}
\leq
w(A)+\frac{\epsilon^{2}w(A)}{9\,L^{2}}+\frac{\epsilon^{2}(1+\epsilon/(3L))}{6L}\,.
\end{equation}
We may assume without loss of generality that $L\leq -1$ also
satisfies $L<-(1+2w(A))/3$. Then the right-hand side of
(\ref{eq:ocho}) is a constant strictly less than $w(A)$, which
contradicts (\ref{eq:uno}). This concludes the proof of
(\ref{eqasympUn}) for $\alpha_{N,k,\mu}$.

If $k$ satisfies the maximum principle, we know by Fuglede's
result (see paragraph after Definition \ref{greedypoints}) and
(\ref{maximumprinciple}) that $U^{\mu}(x)\leq w(A)$ for all $x\in
A$. Therefore the assertions (\ref{asympmk2})-(\ref{eqasympUn})
follow (replacing $S^{*}_{\mu}$ by $A$) for any greedy $k$-energy
sequence $\{\alpha_{N,k}\}$ on $A$ by using the argument presented
above. \hfill $\Box$

\vspace{0.2cm}

\noindent\textbf{Proof of Corollary \ref{co3}.} It is well-known
(see for example \cite{Landkof}) that for any $s<d$ the
equilibrium measure associated with the Riesz kernel $k_{s}$ is
unique and coincides with $\sigma_{d}$. Since
$\supp(\sigma_{d})=S^{d}$, any greedy $k_{s}$-energy sequence
$\{\alpha_{N,s}\}_{N}\subset S^{d}$ is a greedy
$(k_{s},\sigma_{d})$-energy sequence. Therefore by
(\ref{asympmk2}) we obtain that $\{\alpha_{N,s}\}_{N}\in$
AEM$(S^{d};s)$. The values on the right-hand side of
(\ref{asympgreedysld}) are the values of $I_{s}(\sigma_{d})$. The
case $s>0$ follows from formula (1.2) of \cite{KuijSaff} and the
case $s=0$ from formula (2.26) of \cite{Brauchart}. Finally
(\ref{asympdistriesz}) follows from (\ref{asympdistribmk2}).
\hfill $\Box$

\vspace{0.2cm}

\noindent\textbf{Proof of Corollary \ref{corint}.} It is shown in
\cite{Landkof} that for $s<1$ the equilibrium measure associated
with the Riesz kernel $k_{s}$ is
\[
\frac{c_{s}}{(1-x^{2})^{(1-s)/2}}\,dx\,,\qquad x\in(-1,1)\,,
\]
and its energy is given by the value on the right-hand side of
(\ref{asympenergint}). \hfill $\Box$

\vspace{0.2cm}

\noindent\textbf{Proof of Proposition \ref{theonoa1}.} We have
\begin{equation}\label{eq28}
\frac{E_{s}(\alpha_{3\cdot 2^{n},s})-I_{s}(\sigma)(3\cdot
2^{n})^{2}}{(3\cdot 2^{n})^{1+s}}=\frac{1}{3^{1+s}}
\frac{E_{s}(\alpha_{3\cdot
2^{n},s})-I_{s}(\sigma)(2^{n})^{2}-I_{s}(\sigma)2^{2n+3}}
{(2^{n})^{1+s}}\,.
\end{equation}
As will be justified in Section \ref{prueba1} (see Lemma
\ref{lem5}), the relation
\[
E_{s}(\alpha_{3\cdot2^{n},s})=\frac{1}{2}\,\mathcal{E}_{s}(S^{1},2^{n+2})
+\mathcal{E}_{s}(S^{1},2^{n})\,
\]
holds. Therefore, from (\ref{eq28}), it follows that
\[
\frac{E_{s}(\alpha_{3\cdot 2^{n},s})-I_{s}(\sigma)(3\cdot
2^{n})^{2}}{(3\cdot 2^{n})^{1+s}}
\]
\[
=\frac{1}{3^{1+s}}\Big(\frac{\mathcal{E}_{s}(S^{1},2^{n})-I_{s}(\sigma)(2^{n})^{2}}
{(2^{n})^{1+s}}+
\frac{4^{1+s}}{2}\frac{\mathcal{E}_{s}(S^{1},2^{n+2})-I_{s}(\sigma)
(2^{n+2})^{2}}{(2^{n+2})^{1+s}}\Big)\,.
\]
Applying now (\ref{nextordtermminimal}) we get
\[
\lim_{n\rightarrow\infty}\frac{E_{s}(\alpha_{3\cdot
2^{n},s})-I_{s}(\sigma)(3\cdot 2^{n})^{2}}{(3\cdot 2^{n})^{1+s}}=
\Big(\frac{1}{2}\Big(\frac{4}{3}\Big)^{1+s}+\Big(\frac{1}{3}\Big)^{1+s}\Big)
\frac{2\zeta(s)}{(2\pi)^{s}}\,.
\]
Finally, it is easy to check that
$f(s)=\frac{1}{2}(\frac{4}{3})^{1+s}+(\frac{1}{3})^{1+s}<1$ for
all $s\in(0,1)$. \hfill $\Box$

\vspace{0.2cm}

\noindent{\bf Proof of Corollary \ref{conseqsoa}.} Since
$\alpha_{2^{n},s}$ consists of $2^{n}$ equally spaced points (see
Lemma \ref{lem4} below),
$E_{s}(\alpha_{2^{n},s})=\mathcal{E}_{s}(S^{1},2^{n})$, and
therefore
\[
\lim_{n\rightarrow\infty}\frac{E_{s}(\alpha_{2^{n},s})-I_{s}(\sigma)2^{2n}}
{2^{n(1+s)}}=\frac{2\zeta(s)}{(2\pi)^{s}}\,,
\]
but the subsequence $\{\alpha_{3\cdot 2^{n},s}\}_{n}$ provides a
different limit value, given by (\ref{nextorderasymp}). \hfill
$\Box$

\section{{\large Proofs of results from Subsection
\ref{subseccion2}}}\label{prueba1}

In order to prove Proposition \ref{theosm1} we need some auxiliary
lemmas that give a geometric description of greedy $k_{s}$-energy
sequences on $S^{1}$.

\begin{lem}\label{lem2}
Let $s\geq 0$ and consider two points $x_1, x_2 \in S^1$. Set
\[
f(x):=K(|x-x_1|;s)+K(|x-x_2|;s)\,,\qquad x\in S^1\,,
\]
where $K$ is defined in $(\ref{defnRieszK})$. Then on each arc
determined by $x_1$ and $x_2$ the function $f$ has only one
minimum and it is attained at the midpoint of the arc.
\end{lem}
\noindent\textit{Proof.} We write $x_1=e^{i\lambda}$ and
$x_2=e^{i\phi}$, and without loss of generality we assume that
$\lambda=0$ and $\phi\in(0,2\pi)$. We want to show that the
function $g(\theta):=f(e^{i\theta})$ is strictly decreasing on
$(0,\phi/2)$. Since $g(\theta)$ is symmetric on the interval
$(0,\phi)$ with respect to the point $\phi/2$, the location and
uniqueness of the minimum follows. Assume first that $s>0$. We
have that
\[g(\theta)=2^{-\frac{s}{2}}[(1-\cos(\phi-\theta))^{-\frac{s}{2}}+
(1-\cos \theta )^{-\frac{s}{2}}]\,.\] Thus
\[g'(\theta)=\Big(\frac{s}{2}\Big)2^{-\frac{s}{2}}[\sin(\phi-\theta)
(1-\cos(\phi-\theta))^{-\frac{s}{2}-1}-\sin(\theta)(1-\cos(\theta))^{-\frac{s}{2}-1}]\,.\]
Showing that $g'(\theta)<0$ on $(0,\phi/2)$ is equivalent to
\[\frac{\sin(\phi-\theta)}{(1-\cos(\phi-\theta))^{\frac{s}{2}+1}}<\frac{\sin \theta }
{(1-\cos \theta )^{\frac{s}{2}+1}}\,,\quad\theta\in(0,\phi/2)\,.\]
Since $\phi-\theta>\theta$, and the function $(\sin x)/(1-\cos
x)^{\beta}$ is strictly decreasing on $(0,2\pi)$ for $\beta>1$, we
obtain the desired result for $s>0$.

If $s=0$ we have
\[g(\theta)=-\log(2[\cos(\phi/2-\theta)-\cos(\phi/2)])\,,\]
and so the claim is also valid in this case. \hfill $\Box$

\vspace{0.2cm}

\begin{lem}\label{lem4}
Let $s\geq 0$. If $(a_{n})_{n=1}^{\infty}$ is any greedy
$k_{s}$-energy sequence on $S^{1}$, then for every positive
integer $m$, the set $\alpha_{2^{m},s}$ consists of $2^{m}$
equally spaced points, that is,
\[\alpha_{2^{m},s}=\{a_{1}e^{i\frac{2\pi n}{2^{m}}}\}_{n=1}^{2^{m}}\,.\]
\end{lem}
\noindent\textit{Proof.} This property is well-known for $s=0$
(cf.\cite{Baglama}). The following argument applies to all values
of $s\geq 0$. We proceed by induction on $m$. For $m=1$ the result
follows trivially. Assume now that the result is true for $m-1$,
i.e., given any greedy $k_{s}$-energy sequence
$(b_{n})_{n=1}^{\infty}$, the first $2^{m-1}$ points are equally
spaced, and let us show that $\{a_{n}\}_{n=1}^{2^{m}}$ consists of
$2^{m}$ equally spaced points. Consider the function
\[f_{2^{m-1}}(x):=\sum_{n=1}^{2^{m-1}}K(|x-a_{n}|;s),\qquad x\in S^{1}\,.\]
By hypothesis the points $a_{1},\ldots,a_{2^{m-1}}$ are equally
spaced. The symmetry of these points and Lemma $\ref{lem2}$ allow
us to conclude that $f_{2^{m-1}}$ attains its minimum at each
midpoint of the $2^{m-1}$ arcs determined by
$a_{1},\ldots,a_{2^{m-1}}$, and only at these points. Thus,
\begin{equation}\label{eq8}
a_{2^{m-1}+1}\in\{a_{1}e^{i\frac{2\pi(2k-1)}{2^{m}}}\}_{k=1}^{2^{m-1}}\,.
\end{equation}
Now we write
\[f_{2^{m-1}+1}(x)=\sum_{n=1}^{2^{m-1}+1}K(|x-a_{n}|;s)=f_{2^{m-1}}(x)+K(|x-a_{2^{m-1}+1}|;s)\,.\]
The (only) point where the function $f_{2^{m-1}+1}$ attains its
minimum is the point where $K(|x-a_{2^{m-1}+1}|;s)$ attains its
minimum, i.e.,$-a_{2^{m-1}+1}$, since
\[\min_{x\in S^1} f_{2^{m-1}+1}(x)\geq \min_{x\in S^1}f_{2^{m-1}}(x)+\min_{x\in S^1}K(|x-a_{2^{m-1}+1}|;s)\,,\]
and $f_{2^{m-1}}(x)$ and $K(|x-a_{2^{m-1}+1}|;s)$ both attain
their minimum at the same point. In general, by the symmetry of
$\{a_{n}\}_{n=1}^{2^{m-1}}$, if we write
\[f_{2^{m-1}+l}(x)=f_{2^{m-1}}(x)+\sum_{k=1}^{l}K(|x-a_{2^{m-1}+k}|;s)\,\qquad l<2^{m-1}\,,\]
it follows that the point $a_{2^{m-1}+l+1}$ is a point where
$\sum_{k=1}^{l}K(|x-a_{2^{m-1}+k}|;s)$ attains its minimum.
Therefore, the set $\{a_{2^{m-1}+k}\}_{k=1}^{2^{m-1}}$ is formed
by the first $2^{m-1}$ points of some greedy $k_{s}$-energy
sequence. By induction hypothesis,
$\{a_{2^{m-1}+k}\}_{k=1}^{2^{m-1}}$ is formed by $2^{m-1}$ equally
spaced points. From $(\ref{eq8})$ we conclude that
\[\{a_{n}\}_{n=1}^{2^{m}}=\{a_{n}\}_{n=1}^{2^{m-1}}\cup\{a_{2^{m-1}+k}\}_{k=1}^{2^{m-1}}\]
is also formed by equally spaced points. \hfill $\Box$

\vspace{0.2cm}

Two immediate consequences follow from the above proof. The first
one is that greedy $k_{s}$-energy sequences $\{\alpha_{N,s}\}$ on
the unit circle $S^{1}$ are independent of $s$. Hence we will
denote them simply by $\alpha_{N}$. The second consequence is that
the set $\alpha_{3\cdot2^{m}}$ can be written as
\begin{equation}\label{eq9}
\alpha_{3\cdot2^{m}}=S_{2^{m+2}}\setminus S_{2^{m}}\,,
\end{equation}
where $S_{2^{m+2}}$ and $S_{2^{m}}$ are formed, respectively, by
$2^{m+2}$ and $2^{m}$ equally spaced points, and $S_{2^{m}}\subset
S_{2^{m+2}}$.

\vspace{0.2cm}

\begin{lem}\label{lem5}
Let $s\geq 0$. Then given any greedy $k_{s}$-energy sequence
$\{\alpha_{N}\}_{N}$ on $S^{1}$ the following relation holds for
every $n\geq 1$:
\begin{equation}\label{eqrelaminenerg}
E_{s}(\alpha_{3\cdot2^{n}})=\frac{1}{2}\,\mathcal{E}_{s}(S^{1},2^{n+2})
+\mathcal{E}_{s}(S^{1},2^{n})\,.
\end{equation}
\end{lem}
\noindent\textit{Proof.} If $\{x_{k}\}_{k=1}^{N}\subset S^{1}$ is
an arbitrary collection of $N$ equally spaced points, then using
the simple equality
$|e^{i\xi}-e^{i\theta}|=2|\sin(\frac{\xi-\theta}{2})|$, we
conclude that for $s>0$,
\begin{equation}\label{valorminenergcirc}
\mathcal{E}_{s}(S^{1},N)=E_{s}(\{x_{k}\}_{k=1}^{N})=2^{-s}N\sum_{n=1}^{N-1}\sin\Big(\frac{\pi
n}{N}\Big)^{-s}.
\end{equation}
Consider any greedy $k_{s}$-energy sequence
$(\alpha_{N})_{N=1}^{\infty}$ on $S^{1}$. We claim that
\[E_{s}(\alpha_{3\cdot2^{n}})=E_{s}(S_{2^{n+2}})-2^{n+1}\cdot2^{-s}
\sum_{k=1}^{2^{n+2}-1}\sin\Big(\frac{\pi
k}{2^{n+2}}\Big)^{-s}+E_{s}(S_{2^{n}})\,,\] where
$\alpha_{3\cdot2^{n}}=S_{2^{n+2}}\setminus S_{2^{n}}$ is as in
$(\ref{eq9})$. To see this, notice that
$E_{s}(\alpha_{3\cdot2^{n}})$ is obtained by removing twice from
$E_{s}(S_{2^{n+2}})$ all terms $|e^{i\xi}-e^{i\theta}|^{-s}$ where
either $e^{i\xi}\in S_{2^{n}}$ or $e^{i\theta}\in S_{2^{n}}$.

Since
\[E_{s}(S_{2^{n+2}})=\mathcal{E}_{s}(S^{1},2^{n+2})\,,
\qquad E_{s}(S_{2^{n}})=\mathcal{E}_{s}(S^{1},2^{n})\,,\]
$(\ref{eqrelaminenerg})$ follows by applying
$(\ref{valorminenergcirc})$. The case $s=0$ is proved similarly.
\hfill $\Box$

\vspace{0.2cm}

\noindent{\bf Proof of Proposition \ref{theosm1}.} Using
$(\ref{eqrelaminenerg})$ we obtain
\[\frac{E_{s}(\alpha_{3\cdot2^{n}})}{3^{1+s}2^{n(1+s)}}=\frac{1}{3^{1+s}}
\frac{1}{2}\frac{2^{(n+2)(1+s)}}{2^{n(1+s)}}\frac{\mathcal{E}_{s}(S^{1},2^{n+2})}{2^{(n+2)(1+s)}}
+\frac{1}{3^{1+s}}\frac{\mathcal{E}_{s}(S^{1},2^{n})}{2^{n(1+s)}}\,.\]
Simplifying the above expression and applying
$(\ref{asympcircle})$ we conclude that
\[\lim_{n\rightarrow\infty}\frac{E_{s}(\alpha_{3\cdot2^{n}})}{(3\cdot2^{n})^{1+s}}
=\Big(\frac{1}{2}\Big(\frac{4}{3}\Big)^{1+s}+\Big(\frac{1}{3}\Big)^{1+s}\Big)\frac{2\zeta(s)}{(2\pi)^{s}}\,.\]
It is straightforward to check that
$f(s)=\frac{1}{2}\big(\frac{4}{3}\big)^{1+s}+\big(\frac{1}{3}\big)^{1+s}>1$
for all $s>1$. \hfill $\Box$

\vspace{0.2cm}

\noindent{\bf Proof of Proposition \ref{theonoa2}.} First observe
that
\[
\frac{E_{1}(\alpha_{3\cdot 2^{n}})-\frac{1}{\pi}(3\cdot
2^{n})^{2}\log(3\cdot 2^{n})} {(3\cdot 2^{n})^{2}}
\]
\[
=\frac{1}{9}\Big(\frac{(1/2)\,\mathcal{E}_{1}(S^{1},2^{n+2})+\mathcal{E}_{1}(S^{1},2^{n})-
\frac{1}{\pi}(3\cdot 2^{n})^{2}\log(3\cdot 2^{n})}{2^{2n}}\Big)\,.
\]
We add and subtract $(1/\pi)2^{2n}\log(2^{n})$ to obtain
\begin{equation}\label{eqaux1}
\frac{E_{1}(\alpha_{3\cdot 2^{n}})-\frac{1}{\pi}(3\cdot
2^{n})^{2}\log(3\cdot 2^{n})} {(3\cdot 2^{n})^{2}}
\end{equation}
\[
=\frac{1}{9}\Big(\frac{\mathcal{E}_{1}(S^{1},2^{n})-\frac{1}{\pi}2^{2n}\log(2^{n})}{2^{2n}}
+16\frac{(1/2)\,\mathcal{E}_{1}(S^{1},2^{n+2})-\frac{1}{\pi}\Lambda_{n}}{2^{2(n+2)}}\Big)
\]
where $\Lambda_{n}=(3\cdot 2^{n})^{2}\log(3\cdot
2^{n})-2^{2n}\log(2^{n})$. Taking into account that
\[
\Lambda_{n}=\frac{2^{2(n+2)}}{2}\log(2^{n+2})+\log(3)(3\cdot
2^{n})^{2}-8\log(4)2^{2n}
\]
it follows that
\begin{equation}\label{eqaux2}
16\frac{(1/2)\,\mathcal{E}_{1}(S^{1},2^{n+2})-\frac{1}{\pi}\Lambda_{n}}{2^{2(n+2)}}
\end{equation}
\[
=
8\frac{\mathcal{E}_{1}(S^{1},2^{n+2})-\frac{1}{\pi}2^{2(n+2)}\log(2^{n+2})}
{2^{2(n+2)}}+\frac{1}{\pi}(8\log(4)-9\log(3))\,.
\]
Applying (\ref{eqnextordminseq1}), (\ref{eqaux1}) and
(\ref{eqaux2}) we conclude that
\[
\lim_{n\rightarrow\infty}\frac{E_{1}(\alpha_{3\cdot
2^{n}})-\frac{1}{\pi}(3\cdot 2^{n})^{2}\log(3\cdot 2^{n})}
{(3\cdot 2^{n})^{2}}
\]
\[
=\frac{1}{\pi}(\gamma-\log(\pi/2))+\frac{1}{\pi}(\frac{8}{9}\log(4)-\log(3))=
\frac{1}{\pi}(\gamma-\log(\pi/2)+\log(2^{\frac{16}{9}}/3))\,.
\]
\hfill $\Box$

\vspace{0.2cm}

\noindent{\bf Proof of Corollary \ref{conuevo}.} Since
$E_{1}(\alpha_{2^{n}})=\mathcal{E}_{1}(S^{1},2^{n})$ for all $n$,
the result follows from (\ref{eqnextordminseq1}) and
(\ref{eqnextordgreedseq1}). \hfill $\Box$

\section{{\large Proofs of results from subsection
\ref{subseccion4}}}\label{seccion5}

\noindent{\bf Proof of Theorem \ref{theoconjforarcs}.} Assume
first that $\Gamma$ is a Jordan arc. If $x_{1}, x_{2}\in\Gamma$,
we denote by $(x_1,x_2)$ the subarc joining $x_1$ and $x_2$, and
by $l(x_1,x_2)$ its length.

Let $\mathcal{X}_{n}:=\{x_{k,n}\}_{k=0}^{n}$ be a sequence of
configurations on $\Gamma$, where we assume that the points
$x_{k,n}$ are located in successive order. Set
\begin{equation}\label{defndkn}
d_{k,n}:=l(x_{k-1,n},x_{k,n})\,,\qquad k=1,\ldots,n\,.
\end{equation}
In \cite{M-FMRS} the following result was proved:
\begin{theo}\label{theoM-FMRS}
Let $\Gamma$ be a rectifiable Jordan arc in $\mathbb{R}^{p}$. If
$s>1$ and $\{\mathcal{X}_{n}\}_{n}\in \AEM(\Gamma;s)$, then
\begin{equation}\label{asympM-FMRS}
\lim_{n\rightarrow\infty}\sum_{k=1}^{n}\Big|d_{k,n}-\frac{L}{n}\Big|=0\,,\qquad
L:=\mathcal{H}_{1}(\Gamma)\,.
\end{equation}
\end{theo}
We prove Theorem \ref{theoconjforarcs} by contradiction. Let
$\{x_{k}\}_{k=0}^{\infty}\subset\Gamma$ be an arbitrary sequence
of distinct points and set $\mathcal{X}_{n}:=\{x_{k}\}_{k=0}^{n}$.
We will use the notation
$\mathcal{X}_{n}=\{x_{0,n},\ldots,x_{n,n}\}$. Assume that
$\{\mathcal{X}_{n}\}_{n}\in \AEM(\Gamma;s)$. Let $\delta>0$ and
consider the sets
\[
A_{n}^{\delta}:=\{k:
\frac{L-\delta}{n}<d_{k,n}<\frac{L+\delta}{n},\quad 1\leq k\leq
n\}\,,\qquad B_{n}^{\delta}:=\{1,\ldots,n\}\setminus
A_{n}^{\delta}\,.
\]
Let $\epsilon>0$ be a fixed number. Then from (\ref{asympM-FMRS})
there exists $N=N(\epsilon)\in\mathbb{N}$ such that, if $n\geq N$,
\begin{equation}\label{descritical}
\sum_{k=1}^{n}\Big|d_{k,n}-\frac{L}{n}\Big|\leq \epsilon\,.
\end{equation}
If $k\in B_{n}^{\delta}$, then $|d_{k,n}-L/n|\geq \delta/n$, and
from (\ref{descritical}) it follows that
\[
\card(B_{n}^{\delta})\frac{\delta}{n}\leq \epsilon\,,\qquad n\geq
N\,.
\]
Therefore,
\[
\card(A_{n}^{\delta})=n-\card(B_{n}^{\delta})\geq
n\Big(1-\frac{\epsilon}{\delta}\Big)\,,\qquad n\geq N\,.
\]

There are exactly $n$ subarcs $(x_{k-1,n},x_{k,n})$, and when we
add the next $n/2$ points (we may assume that $n$ is even) to the
configuration $\mathcal{X}_{n}$, obviously at most $n/2$ of these
new points will lie in the subarcs $(x_{k-1,n},x_{k,n})$ where
$k\in A_{n}^{\delta}$. Setting
\[
C_{n}^{\delta}:=\{k\in A_{n}^{\delta}: (x_{k-1,n},x_{k,n})\,\,
\mbox{does not contain a new point}\}\,,
\]
we have
\[
\card(C_{n}^{\delta})\geq
n\Big(1-\frac{\epsilon}{\delta}\Big)-\frac{n}{2}=n\Big(\frac{1}{2}-\frac{\epsilon}{\delta}\Big)\,.
\]
Now since the intervals $(x_{k-1,n},x_{k,n})$ with $k\in
C_{n}^{\delta}$ do not contain a new point, there are at least
$\card(C_{n}^{\delta})$ values of $k'$ in $\{1,\ldots,3n/2\}$ such
that $d_{k',3n/2}=d_{k,n}$ for some $k\in C_{n}^{\delta}$. For
these values of $k'$ and the corresponding values of $k$, we have
\[
\Big|d_{k',3n/2}-\frac{L}{3n/2}\Big|=\Big|d_{k,n}-\frac{L}{n}+\frac{L}{3n}\Big|\,.
\]
Now we choose $\delta$ to be any fixed value less than $L/3$, say
$\delta:=L/6$. Then for $k\in C_{n}^{\delta}$,
\[
\Big|d_{k,n}-\frac{L}{n}+\frac{L}{3n}\Big|\geq\Big|\frac{L}{3n}-\Big|\frac{L}{n}-d_{k,n}\Big|\Big|=
\frac{L}{3n}-\Big|\frac{L}{n}-d_{k,n}\Big|>\frac{L}{3n}-\frac{L}{6n}=\frac{L}{6n}\,.
\]
Finally,
\[
\sum_{k'=1}^{3n/2}\Big|\,d_{k',3n/2}-\frac{L}{3n/2}\Big|\geq
n\Big(\frac{1}{2}-\frac{\epsilon}{\delta}\Big)\frac{L}{6n}
=\Big(\frac{1}{2}-\frac{6\,\epsilon}{L}\Big)\frac{L}{6}\,.
\]
But the above estimate contradicts (\ref{descritical}) since we
can select $\epsilon$ sufficiently small so that
\[
\Big(\frac{1}{2}-\frac{6\,\epsilon}{L}\Big)\frac{L}{6}>\epsilon\,.
\]

If $\Gamma$ is a closed Jordan curve, we select an orientation for
it. Then the above reasoning used to prove the result in the case
of Jordan arcs is also applicable. We only have to define
$(x_{k-1,n}, x_{k,n})$ as the subarc joining $x_{k-1,n}$ and
$x_{k,n}$ on which a particle moves from $x_{k-1,n}$ to $x_{k,n}$
following the orientation prescribed. The details of the argument
are left to the reader. \hfill $\Box$

\vspace{0.2cm}

\noindent{\bf Proof of Theorem \ref{theo2}.} We first assume that
$\Gamma$ is a smooth Jordan arc of length $L$. We will reduce the
problem of asymptotics of $\alpha_{N,1}$ on $\Gamma$ to a weighted
problem on $[0,L]$ and then apply Theorem \ref{theoweightset}. Let
$\Phi:[0,L]\longrightarrow\Gamma$ be the natural parametrization
of $\Gamma$ and define $w:[0,L]\times
[0,L]\longrightarrow[0,\infty)$ by
\begin{equation}\label{eq:defnweight}
w(x,y):=\Big|\frac{\Phi(x)-\Phi(y)}{x-y}\Big|^{-1}\,.
\end{equation}
Let $\Psi=\Phi^{-1}$ be the inverse function of $\Phi$. If $a_{n}$
is the $n$-th element of the greedy $k_{1}$-energy sequence on
$\Gamma$, let $b_{n}:=\Psi(a_{n})\in [0,L]$ and
$\beta_{N}:=\{b_1,\ldots, b_{N}\}$. Since for $t=\Phi(x), x\in
[0,L]$,
\[\inf_{t\in \Gamma} \sum_{i=1}^{n-1}\frac{1}{|t-a_{i}|}
=\inf_{x\in [0,L]}\sum_{i=1}^{n-1}\frac{1}{|\Phi(x)-\Phi(b_{i})|}
=\inf_{x\in
[0,L]}\sum_{i=1}^{n-1}\frac{w(x,b_{i})}{|x-b_{i}|}\,,\] it follows
that $\{\beta_{N}\}$ is a greedy $(w,1)$-energy sequence on
$[0,L]$ (see Definition \ref{defngwses}) associated with the
weight function (\ref{eq:defnweight}). Notice that
\[
\mathcal{H}_{1}^{1,w}([0,L])=\int_{0}^{L}w(x,x)^{-1}\,dx=
\int_{0}^{L}|\Phi'(x)|\,dx=L\,.
\]
Applying Theorem \ref{theoweightset} we obtain that
\[\lim_{N\rightarrow\infty}\frac{E_{1}(\alpha_{N,1})}{N^{2}\log N}
=\lim_{N\rightarrow\infty}\frac{E_{1}^{w}(\beta_{N})}{N^{2}\log N}
=\frac{2}{\mathcal{H}_{1}^{1,w}([0,L])}=\frac{2}{L}\,.\]

If $\Gamma$ is a smooth Jordan closed curve and
$\Phi:[0,L]\longrightarrow\Gamma$ is the natural parametrization
of $\Gamma$ ($\Phi(0)=\Phi(L), \Phi'(0)=\Phi'(L)$), we set
\[
\overline{w}(z,\xi):=\frac{|z-\xi|}{|\Phi(x)-\Phi(y)|}\,,\qquad
z=e^{2\pi i x/L}, \xi=e^{2\pi i y/L};\quad x, y\in[0,L]\,,
\]
and apply (with the aid of Theorem \ref{theoweight}) a similar
argument as above on the unit circle $S^{1}$.

In both cases, (\ref{eq:distasymparc}) is a consequence of
(\ref{eq:asymparc}) and Theorem
\ref{theoBHSunweighted}.\hfill$\Box$

\vspace{0.2cm}

\noindent{\bf Proof of Theorem \ref{bestpackrevisited}.} Let $p>1$
be a rational number and let $n\in\mathbb{Z}_{+}$ be such that
$n/p$ is an integer. We denote the first $n+1$ points of the
sequence $\{x_{k}\}_{k=0}^{\infty}$ by
$\mathcal{X}_{n}=\{x_{0,n},\ldots,x_{n,n}\}$, where as in the
proof of Theorem \ref{theoconjforarcs} the points $x_{k,n}$ are
located on $\Gamma$ in successive order. There are exactly $n$
subarcs $(x_{i,n},x_{i+1,n})$. We add to $\mathcal{X}_{n}$ the
next $n/p$ points of the sequence $\{x_{k}\}$. Then there are at
least $(p-1)\,n/p$ subarcs $(x_{i,n},x_{i+1,n})$ not containing a
new point. These subarcs have length at least
$\delta(\mathcal{X}_{n})$. We select $(p-1)\,n/p$ of those.

On the other hand, there are $2\,n/p$ subarcs
$(x_{i,(p+1)n/p},\,x_{i+1,(p+1)n/p})$ remaining with length at
least $\delta(\mathcal{X}_{(p+1)n/p})$. Consequently,
\begin{equation}\label{estpc}
\frac{(p-1)\,n}{p}\,\delta(\mathcal{X}_{n})+\frac{2\,n}{p}\,
\delta(\mathcal{X}_{(p+1)n/p})\leq L\,.
\end{equation}
Thus
\begin{equation}\label{estpc2}
\liminf_{n\rightarrow\infty}n\,\delta(\mathcal{X}_{n})\leq
\frac{p^{2}+p}{p^{2}+2p-1}\,L\,.
\end{equation}
Letting $f(p)$ denote the right-hand side of (\ref{estpc2}), we
see that for $p>1$ the function $f$ attains its minimum when
$p=1+\sqrt{2}$, and
$f(1+\sqrt{2})=\frac{4+3\sqrt{2}}{4+4\sqrt{2}}\,L$, which
establishes (\ref{eqbestpack}).

Let $\mathcal{X}_{n_{k}}$ be a subsequence of configurations such
that
$\lim_{k\rightarrow\infty}n_{k}\,\delta(\mathcal{X}_{n_{k}})=c$.
Notice that we cannot apply (\ref{estpc}) directly because we
cannot assume that $n_{k}/p$ is an integer. Let $\lfloor x
\rfloor$ denote the integral part of $x$ and let $\{x\}:=x-\lfloor
x \rfloor$. Then we get
\begin{equation}\label{estpc3}
\Big(n_{k}-\Big\lfloor\frac{n_{k}}{p}\Big\rfloor\Big)\,\delta(\mathcal{X}_{n_{k}})
+2\Big\lfloor\frac{n_{k}}{p}\Big\rfloor\,\delta(\mathcal{X}_{n_{k}+\lfloor
n_{k}/p\rfloor})\leq L\,.
\end{equation}
Since
\[
\Big|\Big(n_{k}-\Big\lfloor\frac{n_{k}}{p}\Big\rfloor\Big)\,\delta(\mathcal{X}_{n_{k}})-
\frac{(p-1)}{p}\,n_{k}\,\delta(\mathcal{X}_{n_{k}})\Big|=\Big\{\frac{n_{k}}{p}\Big\}
\delta(\mathcal{X}_{n_{k}})\leq \delta(\mathcal{X}_{n_{k}})\,,
\]
it follows that
\begin{equation}\label{estpc4}
\lim_{k\rightarrow\infty}\Big(n_{k}-\Big\lfloor\frac{n_{k}}{p}\Big\rfloor\Big)
\,\delta(\mathcal{X}_{n_{k}})=\frac{(p-1)}{p}\,c\,.
\end{equation}
Similarly,
\[
\Big|(p+1)\Big\lfloor\frac{n_{k}}{p}\Big\rfloor\delta(\mathcal{X}_{n_{k}+\lfloor
n_{k}/p\rfloor})-\Big(n_{k}+\Big\lfloor\frac{n_{k}}{p}\Big\rfloor\Big)
\delta(\mathcal{X}_{n_{k}+\lfloor n_{k}/p\rfloor})\Big|\leq
p\,\delta(\mathcal{X}_{n_{k}+\lfloor n_{k}/p\rfloor})
\]
and thus
\begin{equation}\label{estpc5}
\liminf_{k\rightarrow\infty}\Big(n_{k}+\Big\lfloor\frac{n_{k}}{p}\Big\rfloor\Big)
\delta(\mathcal{X}_{n_{k}+\lfloor
n_{k}/p\rfloor})=\liminf_{k\rightarrow\infty}(p+1)\Big\lfloor\frac{n_{k}}{p}\Big\rfloor\delta(\mathcal{X}_{n_{k}
+\lfloor n_{k}/p\rfloor})\,.
\end{equation}
Since $\liminf_{n\rightarrow\infty}n\,\delta(\mathcal{X}_{n})\leq
\liminf_{k\rightarrow\infty}(n_{k}+\lfloor n_{k}/p \rfloor)
\,\delta(\mathcal{X}_{n_{k}+\lfloor n_{k}/p\rfloor})$, we obtain
from (\ref{estpc3})-(\ref{estpc5}) that
\[
\frac{2}{p+1}\liminf_{n\rightarrow\infty}n\,\delta(\mathcal{X}_{n})\leq
L-\frac{p-1}{p}\,c\,.
\]
Therefore
\[
\liminf_{n\rightarrow\infty}n\,\delta(\mathcal{X}_{n})\leq
g(p):=\Big(1+\frac{1}{p}\Big)\frac{p\,(L-c)+c}{2}\,.
\]
If $c=L$ we get immediately that
$\liminf_{n\rightarrow\infty}n\,\delta(\mathcal{X}_{n})\leq L/2$.
The function $g$ attains a minimum for $p=\sqrt{c/(L-c)}$ and
takes the value $L/2+\sqrt{c\,(L-c)}$ at this point. This proves
(\ref{eqbestpack2}).\hfill $\Box$

\vspace{0.2cm}

\noindent{\bf Proof of Proposition \ref{counterexample}.} Consider
the sequence $\{a_{n}\}_{n=0}^{\infty}\subset[0,1]$ defined as
follows:
\begin{itemize}
\item $a_{0}:=1, a_{1}:=0, a_{2}:=1/2$\,.

\item Assuming that the first $2^{n}+1$ points have been selected,
let $a_{2^{n}+i}:=(2\,i-1)/2^{n+1}$, $1\leq i\leq 2^{n}$.
\end{itemize}

Obviously $\{a_{n}\}_{n=0}^{\infty}$ is a greedy best-packing
sequence on $[0,1]$. However, the sequence of configurations
$S_{N}:=\{a_{n}\}_{n=0}^{N}$ is not uniformly distributed since
\[
\lim_{n\rightarrow\infty}\frac{\card(S_{3\cdot 2^{n-1}}\cap
[0,1/2])}{3\cdot
2^{n-1}+1}=\lim_{n\rightarrow\infty}\frac{2^{n}+1}{3\cdot
2^{n-1}+1}=\frac{2}{3}\neq \frac{1}{2}\,.
\]

Now we consider the sequence
$\{b_{n}\}_{n=1}^{\infty}\subset[0,1]^{2}$ formed in the following
way:
\begin{itemize}
\item[1)] $b_{1}:=(1,1), b_{2}:=(0,0), b_{3}:=(0,1),
b_{4}:=(1,0)$\,.

\vspace{0.1cm}

\item[2)] Assume that the first $(2^{n-1}+1)^{2}$, $n\geq 1$,
points have been selected.

\vspace{0.1cm}

\begin{itemize}
\item[2.1)] We define the next $2^{2(n-1)}$ points as the centers
of the $2^{2(n-1)}$ squares of area $2^{-2(n-1)}$ whose vertices
are the first $(2^{n-1}+1)^{2}$ points
$b_{1},\ldots,b_{(2^{n-1}+1)^{2}}$. These $2^{2(n-1)}$ points are
chosen in an arbitrary order.

\vspace{0.1cm}

\item[2.2)] Now we select the next $2^{n}(2^{n-1}+1)$ points to be
the middle points of the edges of the $2^{2(n-1)}$ squares
mentioned above. The first group of points that we add consists of
those points with abscissa equal to $0$. The second group is
formed by those with abscissa equal to $2^{-n}$. In general, the
points from the $i$-th group have abscissa $(i-1)/2^{n}$. We add
exactly $2^{n}+1$ groups, and in each one of them, the points are
selected in an arbitrary order.
\end{itemize}
\end{itemize}
Figure 1 illustrates the first $221$ points of the sequence
$\{b_{n}\}$.
\begin{figure}[h]
\centering
\includegraphics[totalheight=1.5in,keepaspectratio]{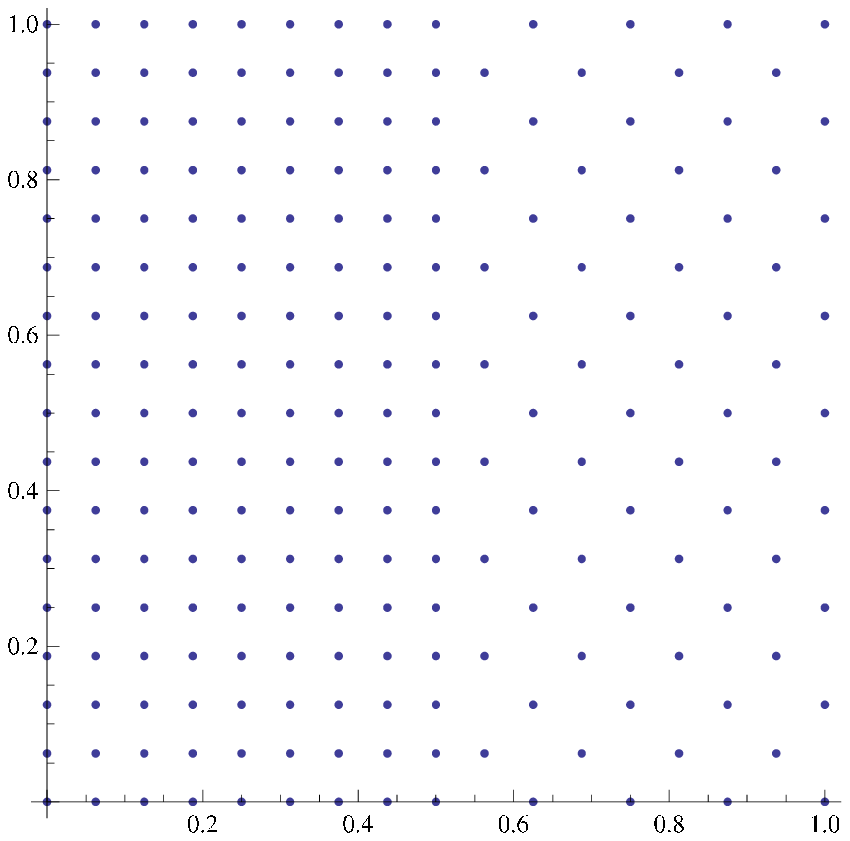}
\label{badsequence} \caption{Greedy best-packing points for
square: a counterexample to a conjecture of Bos.}
\end{figure}

Using Voronoi cell decompositions it is easy to see that
$\{b_{n}\}_{n=1}^{\infty}$ is a greedy best-packing sequence on
$[0,1]^{2}$. To show that the sequence of configurations
$T_{N}:=\{b_{i}\}_{i=1}^{N}$ is not asymptotically uniformly
distributed, we consider the subsequence consisting of
$N(n)=3\cdot 2^{2(n-1)}+7\cdot 2^{n-2}+1$ points. We have that
\[
\lim_{n\rightarrow\infty}\frac{\card(T_{N(n)}\cap [0,1/2]\times
[0,1])}{N(n)}=\lim_{n\rightarrow\infty}\frac{(2^{n-1}+1)(2^{n}+1)}{N(n)}=\frac{2}{3}\neq
\frac{1}{2}\,.
\]
\hfill $\Box$

Using a similar argument it is possible to construct a greedy
best-packing sequence on $[0,1]^{p}\subset\mathbb{R}^{p}$ that is
not asymptotically uniformly distributed.

We remark that it is still plausible that for any infinite compact
$A\subset\mathbb{R}^{p}$ there exists \textit{at least one} greedy
best-packing sequence that is asymptotically uniformly distributed
on $A$.

\section{{\large Proofs of results from subsection \ref{subseccion3}}}

\noindent{\bf Proof of Theorem \ref{theoweight}.} Given a point
$x\in S^d$, we define $C(x,r):=\{y\in S^d: |y-x|\leq r\}$. If
$\sigma_{d}$ denotes the normalized Lebesgue measure on $S^{d}$,
then the following estimates hold (see formulas (3.7) and (3.4) in
\cite{KuijSaff}):
\begin{equation}\label{eq14}
\int_{S^{d}\setminus
C(x,r)}\frac{1}{|x-y|^{d}}d\sigma_{d}(y)=\gamma_{d}\log\Big(\frac{1}{r}\Big)+\mathcal{O}(1)\,,\quad
r\rightarrow 0\,,
\end{equation}
\begin{equation}\label{eq11}
\sigma_{d}(C(x,r))\leq\frac{1}{d}\gamma_{d}\,r^{d}\,,\quad
d\geq2\,,
\end{equation}
where
\begin{equation}\label{defgamma}
\gamma_{d}:=\frac{\Gamma((d+1)/2)}{\Gamma(1/2)\Gamma(d/2)}\,.
\end{equation}
If $d=1$, inequality (\ref{eq11}) is not valid since
$\sigma_{1}(C(x,r))=\frac{2}{\pi}\arcsin(\frac{r}{2})$, but
instead we have
\begin{equation}\label{eq12}
\sigma_{1}(C(x,r))=\gamma_{1}r+\mathcal{O}(r^{3})\,,\quad
r\rightarrow 0\,.
\end{equation}
For $x\in S^{d}$ and $r>0$,
\[\mathcal{H}_{d}^{d,w}(C(x,r))=
\int_{C(x,r)}w(y,y)^{-1}d\mathcal{H}_{d}(y)=\mathcal{H}_{d}(S^{d})\int_{C(x,r)}w(y,y)^{-1}d\sigma_{d}(y)\,.\]
Thus
\begin{equation}\label{eq17}
\mathcal{H}_{d}^{d,w}(C(x,r))\leq
\frac{M\mathcal{H}_{d}(S^{d})\,\gamma_{d}\,r^{d}}{d}\,,\quad d\geq
2\,,
\end{equation}
\begin{equation}\label{eq18}
\mathcal{H}_{1}^{1,w}(C(x,r))\leq
M\mathcal{H}_{1}(S^{1})\gamma_{1}\,r+\mathcal{O}(r^{3})\,,\quad
r\rightarrow 0\,,
\end{equation}
where $M:=\sup\{w(y,y)^{-1}:y\in S^{d}\}$\,.

Let $r\in(0,1)$ be fixed and set
\[D_{i}(r):=S^{d}\setminus C(a_{i},rN^{-\frac{1}{d}})\,,\qquad
D^{N}(r):=\bigcap_{i=1}^{N}D_{i}(r)\,,\] where $a_{i}$ is the
$i$-th element of the greedy $(w,d)$-energy sequence. From
(\ref{eq17}) and (\ref{eq18}) we obtain that
\begin{equation}\label{eq19}
\mathcal{H}_{d}^{d,w}(D^{N}(r))\geq
\mathcal{H}_{d}^{d,w}(S^{d})-\frac{M\mathcal{H}_{d}(S^{d})\gamma_{d}\,r^{d}}{d}\,,\quad
d\geq 2\,,
\end{equation}
\begin{equation}\label{eq20}
\mathcal{H}_{1}^{1,w}(D^{N}(r))\geq
\mathcal{H}_{1}^{1,w}(S^{1})-M\mathcal{H}_{1}(S^{1})\gamma_{1}\,r+\mathcal{O}\Big(\frac{r^{3}}{N^{2}}\Big)\,,\quad
N\rightarrow\infty\,.
\end{equation}
We may assume that the expressions in the right-hand side of the
above inequalities are positive since we can take $r$ sufficiently
close to $0$ and $N$ sufficiently large (we will eventually let
$r\rightarrow 0$ and $N\rightarrow\infty$).

Let $\epsilon>0$. Since the function $w(x,y)/w(x,x)$ is uniformly
continuous on $S^{d}\times S^{d}$, there exists $\delta>0$ such
that
\[\Big|\frac{w(x,y)}{w(x,x)}-1\Big|<\epsilon\,,\qquad\mbox{for}\quad |x-y|<\delta\,.\]
Consider the function
\begin{equation}\label{defnfUndw}
U_{n,d}^{w}(x):=\sum_{i=1}^{n-1}\frac{w(x,a_{i})}{|x-a_{i}|^{d}}\,,\quad
x\in S^{d},\,\,\quad n\geq 2\,.
\end{equation}
From the definition of a greedy $(w,d)$-energy sequence we know
that $U_{n,d}^{w}(a_{n})\leq U_{n,d}^{w}(x)$ for all $x\in S^{d}$.
Let $2\leq n\leq N$ and assume that $r<\delta$. Then
$C(a_{i},rN^{-\frac{1}{d}})\subset C(a_{i},\delta)$ for all $1\leq
i\leq n-1$ and so
\[\int_{D^{N}(r)}U_{n,d}^{w}(x)\,d\mathcal{H}_{d}^{d,w}(x)
\leq\sum_{i=1}^{n-1}\int_{D_{i}(r)}\frac{w(x,a_{i})}{w(x,x)}\frac{d\mathcal{H}_{d}(x)}{|x-a_{i}|^{d}}\]
\[\leq \sum_{i=1}^{n-1}
\Big(\int_{C(a_{i},\delta)\setminus
C(a_{i},rN^{-\frac{1}{d}})}\frac{1+\epsilon}{|x-a_{i}|^{d}}\,d\mathcal{H}_{d}(x)
+\int_{S^{d}\setminus
C(a_{i},\delta)}\frac{w(x,a_{i})}{w(x,x)}\frac{d\mathcal{H}_{d}(x)}{|x-a_{i}|^{d}}\Big)\]
\[\leq (n-1)\Big((1+\epsilon)\mathcal{H}_{d}(S^{d})\int_{S^{d}\setminus C(a_{i},rN^{-\frac{1}{d}})}
\frac{1}{|x-a_{i}|^{d}}d\sigma_{d}(x)+C(w,\delta)\Big)\,,\] where
$C(w,\delta)$ is some constant depending on $\delta$ and $w$.
Using (\ref{eq14}) it follows that
\begin{equation}\label{eqref}
\int_{D^{N}(r)}U_{n,d}^{w}(x)\,d\mathcal{H}_{d}^{d,w}(x)\leq
(n-1)(1+\epsilon)\mathcal{H}_{d}(S^{d})\Big(\frac{\gamma_{d}}{d}\log
N-\gamma_{d}\log r +\mathcal{O}(1)\Big)\,.
\end{equation}
Therefore,
\[E_{d}^{w}(\alpha_{N,d}^{w})= 2\sum_{n=2}^{N}U_{n,d}^{w}(a_{n})\leq 2\sum_{n=2}^{N}
\frac{1}{\mathcal{H}_{d}^{d,w}(D^{N}(r))}\int_{D^{N}(r)}U_{n,d}^{w}(x)\,d\mathcal{H}_{d}^{d,w}(x)\]
\[\leq\frac{N(N-1)}{\mathcal{H}_{d}^{d,w}(D^{N}(r))}(1+\epsilon)\mathcal{H}_{d}(S^{d})
\Big(\frac{\gamma_{d}}{d}\log N-\gamma_{d}\log
r+\mathcal{O}(1)\Big)\,.\] Consequently, from (\ref{eq19}) and
(\ref{eq20}) we get that for $d\geq 1$,
\[
\limsup_{N\rightarrow\infty}\frac{E_{d}^{w}(\alpha_{N,d}^{w})}{N^{2}\log
N}\leq
\frac{1}{\mathcal{H}_{d}^{d,w}(S^{d})-\frac{M\mathcal{H}_{d}(S^{d})\gamma_{d}
\,r^{d}}{d}}(1+\epsilon)\mathcal{H}_{d}(S^{d})\frac{\gamma_{d}}{d}\,.
\]
After letting $r\rightarrow 0$ and $\epsilon\rightarrow 0$ we
obtain that
\[\limsup_{N\rightarrow\infty}\frac{E_{d}^{w}(\alpha_{N,d}^{w})}
{N^{2}\log N} \leq\frac{\mathcal{H}_{d}(S^{d})\,\gamma_{d}}
{\mathcal{H}_{d}^{d,w}(S^{d})\,d}=
\frac{\vol(B^{d})}{\mathcal{H}_{d}^{d,w}(S^{d})}\,.\] Finally,
since $\mathcal{E}_{d}^{w}(S^{d},N)\leq
E_{d}^{w}(\alpha_{N,d}^{w})$ for all $N$, applying
(\ref{theoBHSeq1}) it follows that
\[\lim_{N\rightarrow\infty}\frac{E_{d}^{w}
(\alpha_{N,d}^{w})}{N^{2}\log N}=\frac{\vol(B^{d})}
{\mathcal{H}_{d}^{d,w}(S^{d})}\,.\] The statement about the
weak-star convergence of the normalized counting measure
associated with $\alpha_{N,d}^{w}$ is also an application of
Theorem \ref{theoBHS}. \hfill $\Box$

\vspace{0.2cm}

\noindent\textbf{Remark:} It is not difficult to see that greedy
$k_{s}$-energy sequences on $S^{d}\subset\mathbb{R}^{d+1}$ satisfy
the following property for any $s\in[0,\infty)$. If
$\{a_{n}\}_{n=1}^{\infty}$ denotes such a sequence, then for each
integer $m\geq 1$, the choice of $a_{2m}$ is unique and
$a_{2m}=-a_{2m-1}$.

It is also easily seen that on $S^{2}$ the configuration formed by
the first six points of any greedy $k_{s}$-energy sequence does
not depend on $s$ and is a rotation of the configuration
$\{(1,0,0),(-1,0,0),(0,1,0),(0,-1,0),(0,0,1),\linebreak[1](0,0,-1)\}$
(cf. \cite{Abey}).

\vspace{0.2cm}

\noindent{\bf Proof of Theorem \ref{theoweightset}.} If
$R:=\diam(A)$ is the diameter of $A$, $r<R$ and $x\in A$, then
\begin{equation}\label{eqcaseseqp}
\int_{A\setminus B(x,r)}\frac{1}{|x-y|^{p}}\,dy\leq
\int_{B(x,R)\setminus
B(x,r)}\frac{1}{|x-y|^{p}}\,dy=\mathcal{H}_{p-1}(S^{p-1})\log(R/r)\,.
\end{equation}
Defining
\[D_{i}(r):=A\setminus B(a_{i},r N^{-\frac{1}{p}})\,,\qquad
D^{N}(r):=\bigcap_{i=1}^{N}D_{i}(r)\,,\] where $a_{i}$ is the
$i$-th element of the greedy $(w,p)$-energy sequence, the proof of
Theorem \ref{theoweight} is applicable here and yields the result.
For instance, using (\ref{eqcaseseqp}) the expression similar to
(\ref{eqref}) is
\begin{equation}\label{eqnuevo2}
\int_{D^{N}(r)}U_{n,p}^{w}(x)d\mathcal{H}_{p}^{p,w}(x) \leq
(n-1)(1+\epsilon)\mathcal{H}_{p-1}(S^{p-1})\Big(\frac{1}{p}\log N-
\log r+\mathcal{O}(1)\Big)\,.
\end{equation}
Since $\vol(B^{p})=p^{-1}\mathcal{H}_{p-1}(S^{p-1})$,
(\ref{eqnuevo}) follows from (\ref{eqnuevo2}) and Theorem
\ref{theoBHS}. The limit (\ref{eqnuevo3}) is a consequence of
(\ref{eqnuevo}) and Theorem \ref{theoBHS}.\hfill $\Box$

\vspace{0.2cm}

\noindent{\bf Proof of Theorem \ref{ordergrowth}.} We follow
closely the argument on page 20 of \cite{BHS}. The following
result is known as Frostman's lemma (see \cite{Mattila}).

\begin{lem}\label{frostmanlem}
Let $\delta>0$ and $A$ be a Borel set in $\mathbb{R}^{p}$. Then
$\mathcal{H}_{\delta}(A)>0$ if and only if there exists
$\mu\in\mathcal{M}^{+}(A)$ such that $\mu(A)>0$ and
\begin{equation}\label{frostman}
\mu(B(x,r))\leq r^{\delta},\qquad x\in \mathbb{R}^{p},\quad r>0\,,
\end{equation}
where $B(x,r)$ denotes the open ball centered at $x$ and radius
$r$. Furthermore, one can select $\mu$ so that $\mu(A)\geq
c_{p,\delta}\,\mathcal{H}_{\delta}^{\infty}(A)$, where
$c_{p,\delta}$ is independent of $A$.
\end{lem}

Let $\mu$ be a measure from Lemma \ref{frostmanlem}, and set
$r_{0}:=(\mu(A)/\,2N)^{1/\delta}$. Define the sets
\[
D_{j}:=B(a_{j},r_{0})\,,\qquad
\mathcal{D}_{N}:=A\setminus\bigcup_{j=1}^{N-1}D_{j}\,,
\]
where $a_{j}$ denotes the $j$-th element of the greedy
$(w,s)$-energy sequence. Then, using (\ref{frostman}),
\begin{equation}\label{eqaux33}
\mu(\mathcal{D}_{N})\geq \mu(A)-\sum_{j=1}^{N-1}\mu(D_{j}) \geq
\mu(A)-(N-1)r_{0}^{\delta}>\frac{\mu(A)}{2}>0\,.
\end{equation}
Consider the function $U_{N,s}^{w}$ defined in (\ref{defnfUndw}).
From (\ref{eqaux33}) we obtain
\[
U_{N,s}^{w}(a_{N})\leq \frac{1}{\mu(\mathcal{D}_{N})}
\int_{\mathcal{D}_{N}}U_{N,s}^{w}(x)d\mu(x)\leq \frac{2}{\mu(A)}
\sum_{j=1}^{N-1}\int_{\mathcal{D}_{N}}\frac{w(x,a_{j})}{|x-a_{j}|^{s}}
\,d\mu(x)\]
\[
\leq\frac{2\|w\|}{\mu(A)}\sum_{j=1}^{N-1} \int_{A\setminus
D_{j}}\frac{1}{|x-a_{j}|^{s}}\,d\mu(x)\,,
\]
where $\|w\|:=\sup\{w(x,y):x,y\in A\}$. Set $R:=\diam(A)$. Then
$\mu(A)\leq R^{\delta}$ by (\ref{frostman}). If $y\in A$ and
$r\in(0,R]$, then
\[
\int_{A\setminus B(y,r)}\frac{1}{|x-y|^{s}}\,d\mu(x)\leq
\int_{0}^{r^{-s}}\mu(\{x\in A: \frac{1}{|x-y|^{s}}>t\})dt
\]
\[
\leq
\frac{\mu(A)}{R^{s}}+\int_{R^{-s}}^{r^{-s}}\mu(B(y,t^{-1/s}))dt
\leq R^{\delta-s}+\int_{R^{-s}}^{r^{-s}}t^{-\delta/s}dt
\]
\[
\leq \left\{
\begin{array}{ccc}
R^{\delta-s}+\frac{s}{s-\delta}r^{\delta-s},
& \mbox{if} & s>\delta\,,\\

\\

1+\delta\log\Big(\frac{R}{r}\Big), & \mbox{if} & s=\delta\,.
\end{array}
\right.
\]
Therefore, for $s>\delta$ we obtain
\begin{equation}\label{primera}
U_{N,s}^{w}(a_{N})\leq \frac{2\|w\|}{\mu(A)}(N-1)\Big(R^{\delta-s}
+\frac{s}{s-\delta}r_{0}^{1-s/\delta}\Big)\leq
C_{1}\|w\|\Big(\frac{N}{\mu(A)}\Big)^{s/\delta}\,,
\end{equation}
where $C_{1}>0$ is a constant independent of $N$ and $w$. If
$s=\delta$, then
\begin{equation}\label{segunda}
U_{N,\delta}^{w}(a_{N})\leq \frac{2\|w\|}{\mu(A)}(N-1)\Big(1
+\delta\log\Big(\frac{R}{r_{0}}\Big)\Big)\leq
C_{2}\|w\|\Big(\frac{N\log N}{\mu(A)}\Big)\,,
\end{equation}
where $C_{2}>0$ is also independent of $N$ and $w$. The sequence
$\{U_{i,s}^{w}(a_{i})\}_{N}$ is non-decreasing since
\[
U_{i+1,s}^{w}(a_{i+1})\geq
U_{i,s}^{w}(a_{i})+\frac{w(a_{i+1},a_{i})}{|a_{i+1}-a_{i}|^{s}}\,,\qquad
i\geq 1\,.
\]
Therefore, applying $\mu(A)\geq
c_{p,\delta}\,\mathcal{H}_{\delta}^{\infty}(A)$ and
(\ref{primera})-(\ref{segunda}), Theorem \ref{ordergrowth} readily
follows from
\[
E_{s}^{w}(\alpha_{N,s}^{w})=2\sum_{i=2}^{N}U_{i,s}^{w}(a_{i})\,.
\]
\hfill $\Box$

\vspace{0.2cm}

\noindent{\bf Proof of Corollary \ref{coimp}.} Since
$E_{s}^{w}(\alpha_{N,s}^{w})\geq\mathcal{E}_{s}^{w}(A,N)$ for
every $N$ and $s\geq d$, the lower bounds in (\ref{eqprimera}) and
(\ref{eqsegunda}) follow from (\ref{theoBHSeq3}) and
(\ref{theoBHSeq1}), respectively. The upper bounds follow from
Theorem \ref{ordergrowth}. \hfill $\Box$

\vspace{0.2cm}

\noindent{\bf Proof of Corollary \ref{coimp2}.} Assume the
existence of a point $a\in A$ and $\epsilon>0$ such that
$\{a_{n}\}_{n=1}^{\infty}\cap B(a,\epsilon)=\emptyset$. Let
$\alpha_{N,s}^{w}=\{a_{1},\ldots,a_{N}\}$. Then
\[
E_{s}^{w}(\alpha_{N,s}^{w})=2\,\sum_{1\leq i<j\leq
N}\frac{w(a_{i},a_{j})}{|a_{i}-a_{j}|^{s}} \leq
2\,\sum_{j=2}^{N}\sum_{i=1}^{j-1}\frac{w(a_{i},x)}{|a_{i}-x|^{s}}\,,\]
where the last inequality is valid for any $x\in A$. In
particular, taking $x=a$ we get
\[
E_{s}^{w}(\alpha_{N,s}^{w})\leq\frac{\|w\|}{\epsilon^{s}}N(N-1)\,,
\]
where $\|w\|=\sup\{w(x,y):x,y\in A\}$. This inequality contradicts
the first inequalities in (\ref{eqprimera}) and
(\ref{eqsegunda}).\hfill $\Box$

\vspace{0.2cm}

\noindent{\bf Proof of Proposition \ref{theodensint}.} Assume that
there exists a subinterval $I=[c,d]\subset[a,b]$ for which
(\ref{eq36}) is not satisfied. Let $N_{l}$ be a subsequence such
that
\[
\lim_{l\rightarrow\infty}\frac{(\card\{1\leq n\leq N_{l}: a_{n}\in
I\})^{1+\frac{1}{s}}}{N_{l}}=0\,.
\]
Select $\epsilon>0$ sufficiently small so that
$J=[c+\epsilon/2,d-\epsilon/2]\subset I$ is not empty. If we
define $\nu_{l}:=\card\{1\leq n\leq N_{l}: a_{n}\in J\}$, then
there exists a subinterval of $J$ of length at least
$(d-c-\epsilon)/(\nu_{l}+1)$ not containing any point from
$\{a_{n}\in J:1\leq n\leq N_{l}\}$. Let $x_{l}$ be the center of
such a subinterval. We have, for
$\alpha_{N_{l},s}^{w}=\{a_{1},\ldots,a_{N_{l}}\}$,
\begin{equation}\label{eq38}
E_{s}^{w}(\alpha_{N_{l},s}^{w})=2\sum_{n=2}^{N_{l}}U_{n,s}^{w}(a_{n})\leq
2\sum_{n=2}^{N_{l}}U_{n,s}^{w}(x_{l})=2\sum_{n=2}^{N_{l}}
\sum_{i=1}^{n-1}\frac{w(x_{l},a_{i})}{|x_{l}-a_{i}|^{s}}
\end{equation}
\[
\leq 2\|w\|\Big[\frac{N_{l}-1}{|x_{l}-a_{1}|^{s}}+\frac{N_{l}-2}
{|x_{l}-a_{2}|^{s}}+\cdots+\frac{1}{|x_{l}-a_{N_{l}-1}|^{s}}\Big]
=2\|w\|(S_{I,l}+T_{I,l})\,,
\]
where $\|w\|=\sup\{w(x,y):x,y\in [a,b]\}$ and
\[
S_{I,l}:=\sum_{a_{i}\in I,\,1\leq i\leq
N_{l}-1}\frac{N_{l}-i}{|x_{l}-a_{i}|^{s}}\,,\qquad
T_{I,l}:=\sum_{a_{i}\notin I,\,1\leq i\leq
N_{l}-1}\frac{N_{l}-i}{|x_{l}-a_{i}|^{s}}\,.
\]

For each $a_{i}\notin I$, $|a_{i}-x_{l}|\geq \epsilon/2$; hence
\begin{equation}\label{eq37}
2\,T_{I,l}\leq (2/\epsilon)^{s}\,N_{l}^{2}\,.
\end{equation}
If $a_{i}\in I, 1\leq i\leq N_{l}-1,$ then $|a_{i}-x_{l}|\geq
(d-c-\epsilon)/2(\nu_{l}+1)$. Therefore, if we define
$\tau_{l}:=\card\{1\leq i\leq N_{l}-1: a_{i}\in I\}$, it follows
that
\begin{equation}\label{eq39}
2\,S_{I,l}\leq\frac{2^{s+1}}{(d-c-\epsilon)^{s}}\,(\nu_{l}+1)^{s}\,\tau_{l}\,N_{l}\,.
\end{equation}
By hypothesis, $\tau_{l}^{1+s}/N_{l}^{s}\rightarrow 0$ as
$l\rightarrow\infty$. We deduce from (\ref{eq38})-(\ref{eq39})
that
\[
\lim_{l\rightarrow\infty}\frac{E_{s}^{w}(\alpha_{N_{l},s}^{w})}{N_{l}^{1+s}}=0\,,
\]
which contradicts the fact that
\[
\liminf_{N\rightarrow\infty}\frac{E_{s}^{w}(\alpha_{N,s}^{w})}{N^{1+s}}\geq
\lim_{N\rightarrow\infty}\frac{\mathcal{E}_{s}^{w}([a,b],N)}{N^{1+s}}=\frac{2\zeta(s)}
{\mathcal{H}_{1}^{s,w}([a,b])^{s}}>0\,.
\]
\hfill $\Box$

\vspace{0.4cm}

\noindent{\bf Acknowledgments.} The authors wish to thank
Professors A. Aptekarev and D. Hardin for many valuable
discussions with us about the content of this paper.

\vspace{0.5cm}

Center for Constructive Approximation, Department of Mathematics,
Vanderbilt University, Nashville, TN 37240, USA
$<$abey.lopez@vanderbilt.edu$>$.

\vspace{0.2cm}

Center for Constructive Approximation, Department of Mathematics,
Vanderbilt University, Nashville, TN 37240, USA
$<$edward.b.saff@vanderbilt.edu$>$.

\end{document}